\documentclass[aps,pre,preprint,superscriptaddress,longbibliography]{revtex4-1}
\usepackage{latexsym}
\usepackage{amsmath}
\usepackage{amsfonts}
\usepackage{amssymb}
\usepackage[utf8]{inputenc}
\usepackage[T1]{fontenc}
\usepackage[dvipdfm]{color,graphicx}
\usepackage{verbatim}
\usepackage{t1enc}
\usepackage{mathrsfs}
\usepackage{anysize}
\usepackage{tocbibind}
\settocbibname{REFERENCES}
\usepackage[percent]{overpic}
\usepackage{float} 
\usepackage{appendix}
\usepackage{booktabs} 
\usepackage{natbib}
\usepackage{array}
\usepackage{xcolor}
\usepackage[normalem]{ulem}
\usepackage[colorlinks=true,linkcolor=blue,citecolor=blue,urlcolor=blue, bookmarks=false]{hyperref}

\newcommand{\panelgraphics}[4]{%
  \begin{minipage}[t]{#1\linewidth}
    \centering
    \textbf{#2}\\[-0.4ex]
    \includegraphics[width=\linewidth,alt={#4}]{#3}
  \end{minipage}%
}

\begin{document}
\title{Household Bubbling Strategies for Epidemic Control and Social Connectivity}

%Steven Ruggles, Lara Cleveland, Rodrigo Lovaton, Sula Sarkar, Matthew Sobek, Derek Burk, Dan Ehrlich, Quinn Heimann, Jane Lee, and Nate Merrill. Integrated Public Use Microdata Series, International: Version 7.6 [dataset]. Minneapolis, MN: IPUMS, 2025.
%https://doi.org/10.18128/D020.V7.6

\author{L.  D. Valdez}

\affiliation{Departamento de F\'isica, FCEyN, Universidad Nacional de Mar del Plata, Mar del Plata 7600, Argentina.}
\affiliation{Instituto de Investigaciones F\'isicas de Mar del Plata (IFIMAR), CONICET, Mar del Plata 7600, Argentina.}

\author{J. H. Peressutti}

\affiliation{Departamento de F\'isica, FCEyN, Universidad Nacional de Mar del Plata, Mar del Plata 7600, Argentina.}
\affiliation{Instituto de Investigaciones F\'isicas de Mar del Plata (IFIMAR), CONICET, Mar del Plata 7600, Argentina.}
  \date{\today}

\begin{abstract}

During the COVID-19 crisis, policymakers have implemented social bubble merging strategies, which allowed people from different households to meet and interact. Although these measures can mitigate the negative effects of extreme isolation, they also introduce additional contacts that may facilitate disease spread. As a result, several modeling studies have explored the epidemiological impact of different household-merging strategies, in which the selection of households to be merged is guided by specific demographic criteria, such as household size or the age composition of their members. Here, we investigate an alternative pairing strategy in which households are merged according to the number of economically active (working) members. We develop a mathematical model of household networks using real demographic data from multiple regions around the world, and simulate a lockdown scenario in which only economically active individuals can leave their households, while the remaining non-working members stay indoors. By using numerical simulations and the generating function technique, we then estimate the epidemic risk for different household merging strategies. We find that merging strategies based on the number of working members can keep epidemic risk at similar levels as those based on household size. Moreover, the worker-based approach allows significantly more people to form larger social bubbles, exceeding 40\% of the population in some countries. We find that merging households with at most one worker provides the best balance between controlling epidemic risk and addressing people's need for social contact.
\end{abstract}

\maketitle

\section{Introduction}
Social relationships are a fundamental human need that cannot be suppressed indefinitely without consequences. A large number of studies have shown that isolation and loneliness are associated with poorer physical and mental health, and, in the most severe cases, with increased mortality rates~\cite{beller2018loneliness,lauder2006comparison}. These consequences became especially clear during the COVID-19 lockdowns, which have been called the "world's biggest psychological experiment"~\cite{van2020lockdown}. When the COVID-19 pandemic emerged, many countries imposed very restrictive stay-at-home orders and other non-pharmaceutical interventions to slow down the spread of the disease~\cite{nader2021onset}. While most of the world's population initially remained confined to their homes with reasonable compliance, over time, social isolation and loneliness gradually eroded the well-being of the population~\cite{farrell2023loneliness,morgan2024older}. At the same time, more people began to leave their homes without permission, often due to emotional fatigue or economic necessity, increasing the risk of community transmission, as well as the tension between citizens and law enforcement agencies~\cite{shearston2021social,gibson2020businesses,shodunke2022enforcement}.

During the COVID-19 crisis, one strategy that gained popularity in several countries to mitigate the negative effects of extreme isolation was the so-called {\it social bubble} strategy. A social bubble is a metaphorical term that refers to a small group of people who are free to interact with each other but have little or no contact with those outside the group~\cite{long2020living,leng2021effectiveness,trnka2020blowing}. The main idea of this strategy is that people are confined to their bubbles for a certain period in order to contain the spread of infection.  However, and more importantly, the size of these social bubbles can later be adjusted dynamically (for example, by letting two households to join together) in order to balance disease transmission control with people's need for social contact. An illustrative example of the flexibility of these bubbles can be seen in the case of New Zealand~\cite{long2020living,kearns2021big,trotter2021ways}. At the beginning of the COVID-19 pandemic, this country restricted bubbles to individual households as a precautionary measure. Combined with other non-pharmaceutical interventions, this approach helped to quickly contain the disease. Then, as the number of cases declined and stabilized, the government allowed household bubbles to merge in pairs in order to alleviate the consequences of extreme isolation. Similarly, in Nova Scotia (Canada), two months after the COVID-19 pandemic was declared, health agencies there also permitted households to join in pairs~\cite{quon2020nova_scotia_coronavirus}. A later study found that this bubble approach helped mothers feel less stressed and improved their sense of social support~\cite{humble2024mothers}.

Beyond the psychological benefits of bubble strategies documented in the literature, researchers have recently developed mathematical and computational models to estimate how effective these strategies are in controlling disease spread. For instance, Hill proposed an agent-based model to simulate the spread of the virus in a synthetic population of households under different bubble policies, and the results of this study indicated that smaller groups can reduce the risk of infection more effectively than larger or longer-lasting groups~\cite{hill2023modelling}. In a different direction, Danon et al.~\cite{danon2021household} proposed a network model in which each household was represented as a clique (i.e., a complete subgraph), and used the percolation approach~\cite{stauffer2018introduction,newman2001random,newman2002spread} to estimate the risk of an epidemic. They explored multiple strategies for merging cliques/households and found that one of the best scenarios is to join single-person households with another household of any size. On the other hand, Leng et al.~\cite{leng2021effectiveness} explored multiple merging criteria, including strategies that allow only households with young children to pair up, and strategies that allow single-person households to be merged.

While previous modeling studies have used household size as one of the criteria for designing household bubble strategies~\cite{danon2021household,leng2021effectiveness}, the number of household members who are economically active or otherwise exposed to the outside world may be a relevant factor for designing more flexible and socially beneficial merging strategies. As a simple hypothetical example, consider two households: a single person living alone and a household with four members. At first glance, when health authorities have to decide which households can merge to form larger bubbles, it might seem riskier to include this four-person household in a larger bubble. This is because larger households often have more connections to the outside world, which could increase community transmission. However, this is not always the case.  Imagine this four-person household consists of two parents and two children, where only one parent goes to work while the others stay home and do not interact with anyone outside the family. If the working parent becomes infected, this person could spread the disease to the other family members. But since the other three are isolated at home, they will not spread the disease to the rest of the population. From an epidemic perspective, this household poses the same community risk as a single worker living alone. Therefore, if we applied a bubble strategy, this four-person household should have the same epidemic impact as a single-person household. Moreover, allowing the four-person household to merge with another home to form a larger bubble would clearly provide psychological benefits to more people. This hypothetical example shows that household size should not be the only factor when deciding which households can merge.

In this paper, we introduce an epidemic model that explicitly accounts for household structure, labor activity, and social bubble formation. We use a network-based model where households are represented as fully connected subgraphs  (i.e., cliques), and only household members who are economically active are allowed to establish connections with other people outside their households. These household networks are built using real demographic data from Argentina, China, Israel, and Spain, and on top of these networks, we simulate a classic susceptible-infected-recovered epidemic process. Based on this model, we then explore various social bubble strategies by merging households according to the number of working members.  By using the generating function technique, we explore how epidemic risk changes under different merging scenarios. In particular, we ask whether merging criteria based on the number of economically active members can improve the balance between reducing social isolation and limiting epidemic risk, compared with strategies based only on household size.

The remainder of this paper is organized as follows. In Sec.~\ref{sec.modeltheor}, we describe the household network, the epidemic dynamics, and the household merging strategies considered in this study. In Sec.~\ref{sec.ResulNoMerg}, we present baseline results for the model without household merging and derive analytical expressions for the epidemic threshold. We then analyze the impact of different bubble strategies on the epidemic threshold in Sec.~\ref{sec.ResulMerg}. Finally, we summarize our main findings in the concluding section.

\section{Model}\label{sec.modeltheor}
In the following subsections, we will describe the household network used in our model, the epidemic dynamics, and the rules of our bubbling strategy. The main algorithms corresponding to the model described in this section are publicly available in our GitHub repository~\cite{gith01}.

\subsection{The household network}\label{sec.houseNet}

First, let us introduce some definitions and notation. We consider a population consisting of $H$ households. Each household is modeled as a clique, that is, a complete subgraph where all household members are connected by pairwise links. Treating small groups as cliques is a widely used approach in network models of disease spread~\cite{danon2021household,valdez2023epidemic,valdez2024explosive,valdez2025superexponential,rizi2024effectiveness,rizi2025homophily,ma2013effective,volz2011effects}. Other, more detailed descriptions are also possible. For example, hypergraphs and simplicial complexes can represent higher-order or group interactions,  in which the probability of infection depends on the collective state of the group rather than on pairwise links between its individual members~\cite{sun2025optimal,luo2025optimal,guo2024pattern,yan2026hyperedge,battiston2020networks,boccaletti2023structure,wang2024epidemic}. In the present work, however, we will focus only on the clique representation.

We denote the number of members/nodes in a clique as $s$ (also referred to as the clique size). For a clique with $s$ members, we assume that a number $w$ of them (with $w\leq s$) are economically active individuals, or workers. Non-workers maintain only internal (household) connections, while workers form connections with individuals outside their household. The number of external connections that each worker has is called the external degree and is denoted as $k_E$. For simplicity, we assume that workers connect randomly with one another and these external connections remain static over time. 

\begin{figure}[htbp]
\centering
\panelgraphics{0.48}{(a)}{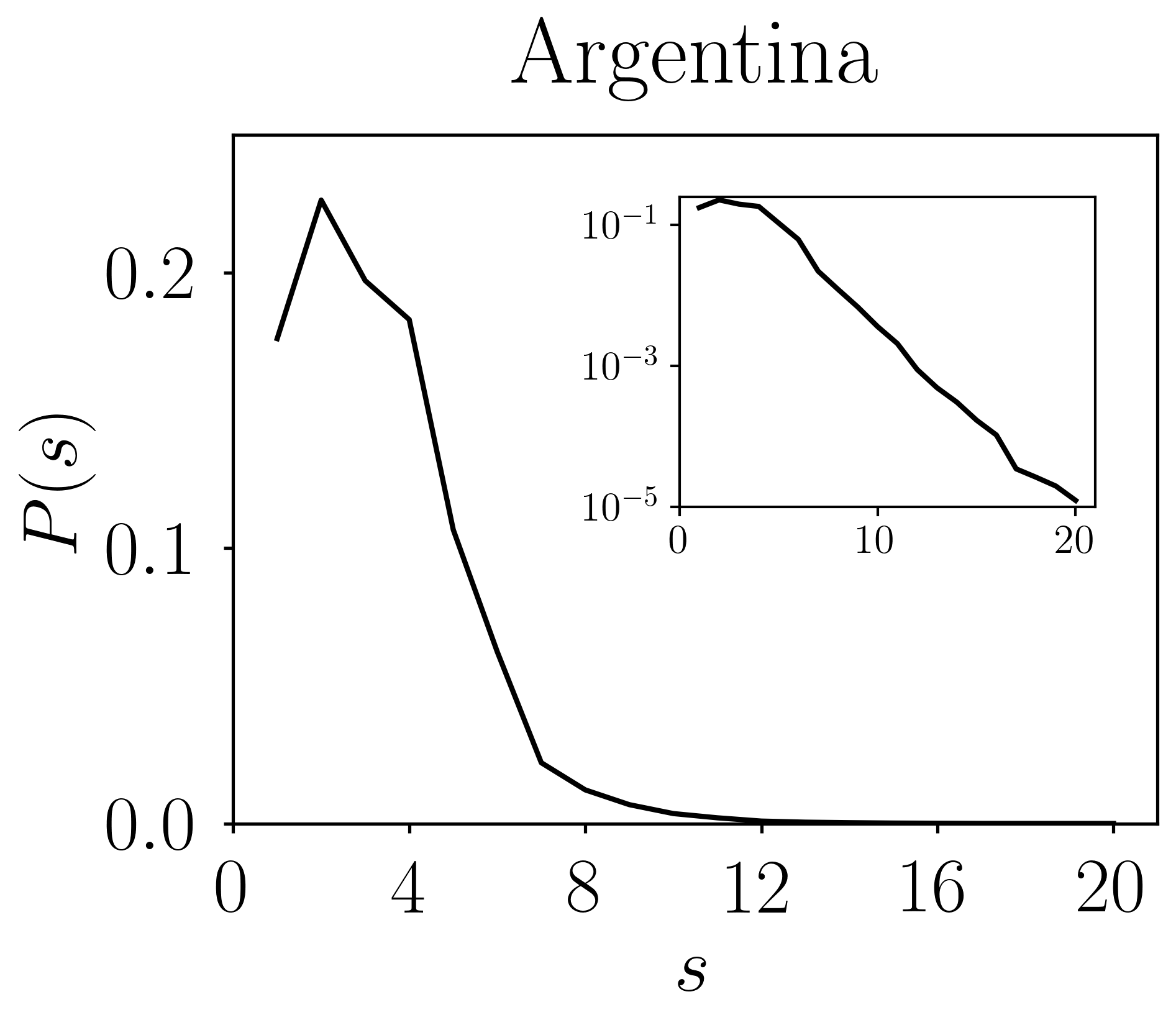}
{Household size distribution for Argentina.}
\hfill
\panelgraphics{0.48}{(b)}{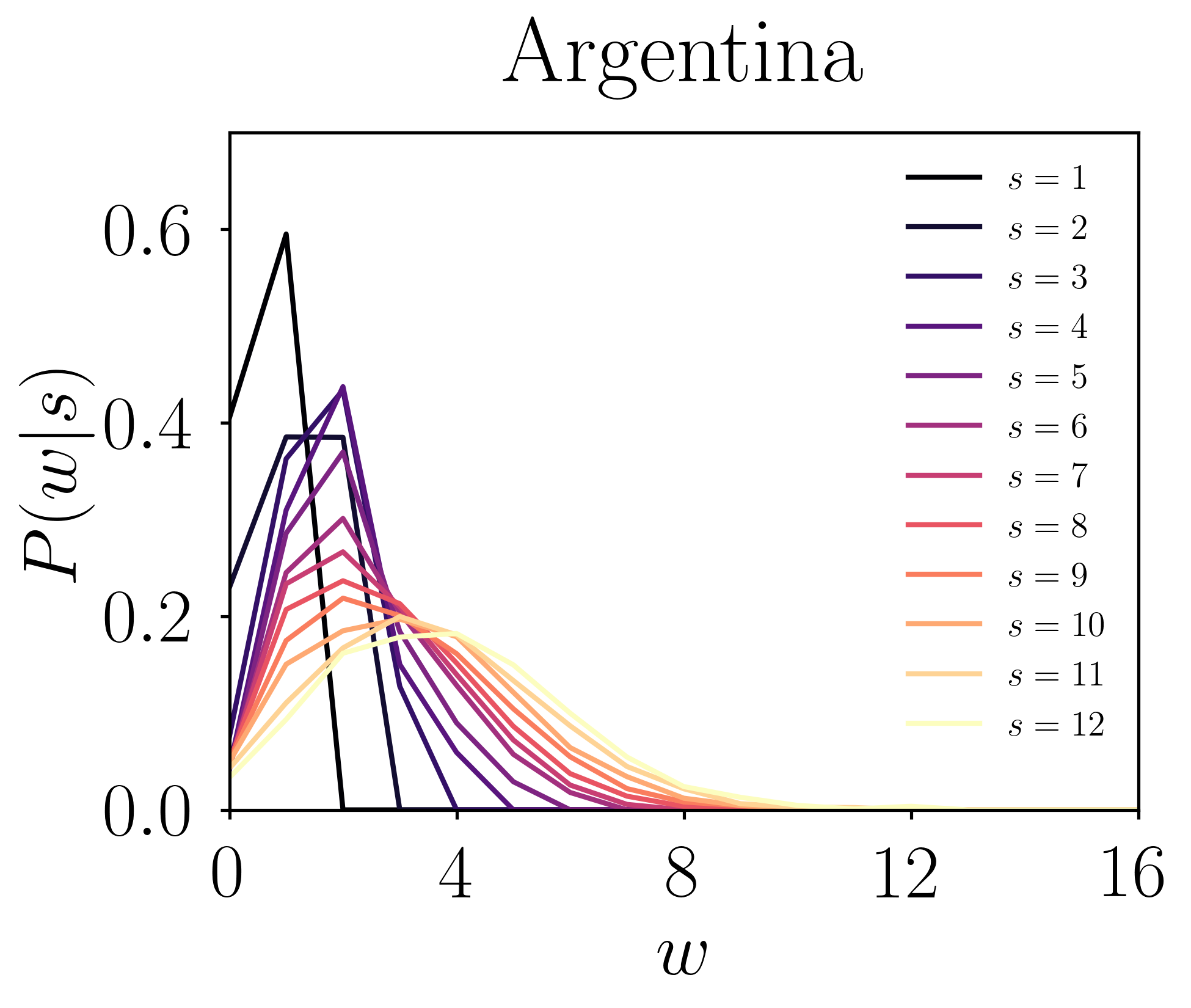}
{Conditional distribution of the number of workers within households in Argentina.}
\caption{Panel a) In the main figure, we show the household size distribution $P(s)$ for Argentina, and in the inset, we display the same curve but on a log-linear scale. Panel b) Probability of having $w$ workers in households of size $s$ (ranging from $s=1$ to $s=12$) in Argentina.}
\label{fig.PsPwsArg}
\end{figure}

Based on these definitions, we introduce the following functions:
\begin{itemize}
\item $P(s)$ is the probability that a randomly selected clique has size $s$.
\item $P(w,s)$ is the joint probability that a household has size $s$ and contains $w$ workers (with $0\leq w\leq s$). Note that $P(s)=\sum_{w=0}^sP(w,s)$. In this work, all economically active individuals are treated identically, and no distinction is made between essential and non-essential workers.
\item $P(w|s)$ is the conditional probability of having $w$ workers in a clique of size $s$. This is calculated as $P(w|s)=P(w,s)/P(s)$.
\item $P(k_E)$ is the probability that a worker has $k_E$ external connections.
\end{itemize}
In this work, we use realistic distributions for both household size $P(s)$ and the number of workers within households $P(w|s)$ for several countries. These were derived from census micro-data provided by IPUMS International~\cite{ruggles2025integrated}. Figure~\ref{fig.PsPwsArg}a shows $P(s)$ for Argentina, while the corresponding distributions for China, Spain, and Israel are presented in the Supplementary Material. Despite slight variations between these countries, $P(s)$ consistently shows a unimodal shape with a peak around $s=2$ or $s=3$, followed by an approximately exponential decay for larger $s$.

On the other hand, Fig.~\ref{fig.PsPwsArg}b displays the conditional distribution $P(w|s)$ for Argentina. From this figure, we can observe that $P(w|s)$ is unimodal as well, with peaks at $w=1$ or $w=2$ for $s\lesssim 5$, which may correspond to a common family structure of two adults with children. On the other hand, for households with $s>5$, the peak of $P(w|s)$ shifts to higher values of $w$.

Finally, for external connections, we will consider two distributions for worker connectivity $P(k_E)$:
\begin{itemize}
\item truncated Poisson distribution $Pois(\lambda,k_{min},k_{max})$, defined as:
\[
Pois(\lambda,k_{min},k_{max}) = 
\begin{cases} 
      c \frac{\lambda^k\exp(-\lambda)}{k!}, & \text{if } k_{min}\leq k \leq k_{max} \\
      0, & \text{otherwise}
\end{cases}
\]\label{eq.trunPois}
where $c$ is a normalization constant.
\item truncated power-law distribution $PL(\lambda,k_{min},k_{max})$, defined as:
\[
     PL(\lambda,k_{min},k_{max}) = 
     \begin{cases} 
          c k^{-\lambda}, & \text{if } k_{min}\leq k \leq k_{max} \\
          0, & \text{otherwise}
     \end{cases}
    \]\label{eq.trunPLsupp}
    where $c$ is a normalization constant.
\end{itemize}
Networks whose nodes have an external degree following a truncated Poisson distribution will be referred to as ER networks, whereas those following a truncated power-law distribution will be referred to as SF networks.

From the above distributions, one can compute the following first and second moments (which will be useful in the subsequent sections to estimate the critical epidemic threshold): 
\begin{itemize}
    \item for household sizes, we denote the mean and second moment as: $\langle s\rangle=\sum_s s P(s)$ and $\langle s^2\rangle=\sum_s s^2 P(s)$, respectively,
    \item for the number of workers, we denote the first and second moments as: $\langle w\rangle=\sum_s \sum_{w=0}^s w P(w,s)$ and $\langle w^2\rangle=\sum_s \sum_{w=0}^s w^2 P(w,s)$, respectively,
    \item for external connections, we define the first and second moments as: $\langle k_E\rangle=\sum_{k_E} k_E P(k_E)$ and $\langle k_E^2\rangle=\sum_{k_E} k_E^2 P(k_E)$, respectively.  
    We will use $\langle k_E^2\rangle-\langle k_E\rangle$ as an indicator of the heterogeneity of the external degree distribution. Larger values of this quantity indicate a broader distribution and, therefore, more heterogeneous external connections.
\end{itemize}

\subsection{Epidemic model}\label{sec.epidmod}
On top of the household network described above, we will study an epidemic process based on the classic susceptible-infected-recovered (SIR) model~\cite{pastor2015epidemic}. In this model, susceptible individuals are healthy but have no immunity, while infected individuals can transmit the disease to their susceptible contacts. Finally, recovered people have developed immune defenses against the disease and cannot become infected again. Here, we will implement a discrete-time version of this model, in which at each time step, the following two transitions occur: i) all infected individuals transmit the disease to each susceptible neighbor with probability $\beta$, and ii) individuals who have been infected for $t_r$ time steps (the recovery time) transition to the recovered state with probability 1.

In our work, we will extend the classic discrete-time SIR model by considering different transmission probabilities within and across households. We denote by $\beta^I$ the infection probability between infected and susceptible individuals in the same household. On the other hand, we denote by $\beta^E$ the transmission probability outside households. We keep these two probabilities separate instead of using a single effective infection probability  because epidemiological studies showed that transmission risk can differ between household and non-household settings~\cite{bulfone2021outdoor,lynch2022literature,bi2021insights}. Additionally, as we will see in the following sections, our model is analytically tractable in the two limiting cases $\beta^I\to 0$ and $\beta^I=1$.

\subsection{Merging strategy}\label{sec.bubstratt}

\begin{figure}[htbp]
\centering
\includegraphics[width=0.9\linewidth,alt={Schematic representation of two households merging into one larger social bubble.}]{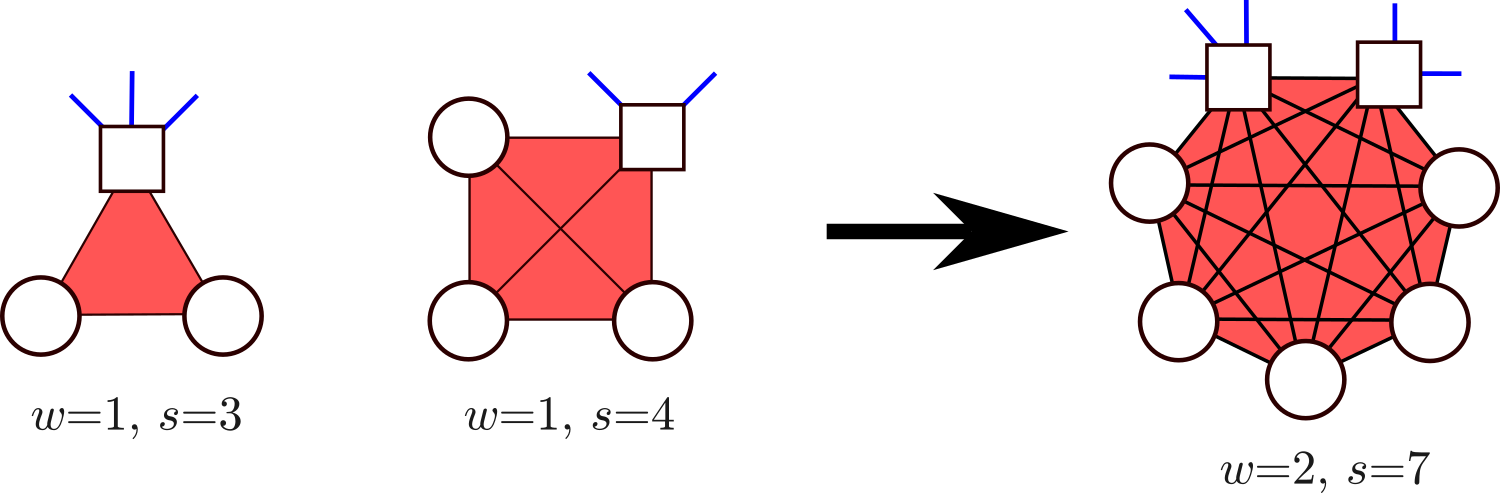}
\caption{Schematic figure showing how two households (cliques) are merged into a single bubble. Circles represent non-workers without external contacts, and squares represent workers with external contacts. Black links indicate internal household connections, while blue links indicate external connections. On the left, a household of three members with one worker ($s=3$, $w=1$) and a household of four members with one worker ($s=4$, $w=1$) are shown separately. On the right, the two households are combined into a single bubble with seven members and two workers ($s=7$, $w=2$).}
\label{fig.bubble}
\end{figure}

Beyond treating individual households as bubbles, we will also consider a set of different bubbling strategies in which these pre-existing households are merged into larger units, as illustrated in Fig.~\ref{fig.bubble}. For simplicity, we assume that merging two households/cliques will lead to a larger fully connected bubble or clique, as shown in that figure.

We will explore the following household merging strategies:
\begin{itemize}
\item Scenario $w^*=1$: every household with at most one worker ($w\leq 1$) merges with another household that also has at most one worker.
\item Scenario $w^*=2$: every household with at most two workers merges with another household that also has at most two workers.
\item Scenario $w^*=w$: every household merges with another household, regardless of the number of workers. 
\item Scenario $1+s$: every single-person household bubble merges with another household of size two or larger. %This scenario was previously studied in Ref.~\cite{danon2021household}.
\item Scenario $2+s$: every household with at most two people merges with another household of size three or larger. %This scenario was also previously studied in Ref.~\cite{danon2021household}.
\end{itemize}
After merging households according to any of these strategies, the probability distribution for the number of workers within households changes from the original $P(w,s)$ to a new distribution which we denote by $\widetilde{P}(w,s)$.

In real life, when one of the bubbling strategies listed above is implemented, one might expect that certain pairing patterns would emerge. For instance, it seems reasonable to expect that a typical family of two children with two parents would form a larger bubble with a household of one or two retired people (the grandparents). However, to keep the model simple, we will assume no correlations exist. For example, for the $w^*=1$ strategy, households are paired in an uncorrelated way, with the only requirement that each household has at most one worker. One advantage of this simplification is that the resulting distribution $\widetilde{P}(w,s)$ can be derived directly from the original distribution $P(w,s)$. The details of this calculation are given in Appendix~\ref{app.calcPtilde}.

For later comparison between different household merging strategies, we also define the average number of internal links per household,
\begin{equation}
\langle \ell_I \rangle = \sum_s \frac{s(s-1)}{2}\, P(s).
\end{equation}
This quantity increases when households are merged into larger bubbles, and we interpret larger values of $\langle \ell_I \rangle$ as indicating greater potential opportunities for social interaction. In line with this interpretation, studies showed that larger social networks and face-to-face communication can reduce social isolation and improve mental health~\cite{liang2024person,domenech2021social}.

It is worth noting that the $1+s$ and $2+s$ scenarios considered here differ slightly from the corresponding strategies studied in Ref.~\cite{danon2021household}. For example, in our formulation, single-person households in the $1+s$ scenario can only merge with households of size two or larger, but not with other single-person households. This contrasts with the $1+s$ strategy studied in Ref.~\cite{danon2021household}, where single-person households were also allowed to merge among themselves. We introduce this restriction to simplify the mathematical derivation of $\widetilde{P}(w,s)$. However, this modeling choice has an important consequence: by preventing small households from pairing together, our merging strategy $1+s$ produces larger bubbles on average than those considered in Ref.~\cite{danon2021household}. This means that the scenarios studied here will provide greater social benefits because more people will see their bubbles grow and gain opportunities for social interaction. At the same time, if our $1+s$ strategy produces only a small increase in epidemic risk despite creating larger bubbles, then it is reasonable to expect that the $1+s$ strategy in Ref.~\cite{danon2021household} (which produces smaller bubbles) would generate an even smaller epidemic impact.

\section{Results}

\subsection{Results with no bubble merging}\label{sec.ResulNoMerg}
In this section, we will present the results of our model for Argentina without household merging as our baseline scenario. The corresponding results for other countries are provided in the Supplementary Material. For the simulations, we use the empirical household size distribution $P(s)$ and the worker distribution $P(w|s)$ shown in Sec.~\ref{sec.houseNet}. Recall that in our model, only workers have external contacts. Here, we will explore two distributions for $P(k_E)$:
\begin{itemize}
\item $Pois(\lambda=2,k_{min}=1,k_{max}=5)$, representing a scenario in which workers have on average about two external contacts. Here, $\langle k_E\rangle\approx 2.23$ and $\langle k_E^2\rangle\approx 6.27$.
\item $PL(\lambda=2.5,k_{min}=1,k_{max}=100)$, representing a scenario in which many workers have very few external contacts, while a smaller number have many. Here, $\langle k_E\rangle\approx 1.80$ and $\langle k_E^2\rangle\approx 13.8$. 
\end{itemize}

Although these two distributions have a similar mean number of external contacts, they differ in their level of heterogeneity.  Following the criterion introduced in Sec.~\ref{sec.houseNet}, we quantify this heterogeneity through the difference $\langle k_E^2\rangle-\langle k_E\rangle$. For the ER case, this difference is $\langle k_E^2\rangle-\langle k_E\rangle\approx 4.04$, whereas for the SF case it is $\langle k_E^2\rangle-\langle k_E\rangle\approx 12.0$. Therefore, the ER network has lower degree heterogeneity, while the SF network has higher degree heterogeneity. In this sense, we will refer to the ER case as having more homogeneous external connections and to the SF case as having more heterogeneous external connections. 

For our numerical simulations, we generate random networks with cliques using a variant of the configuration model (see Ref.~\cite{molloy1995critical}), and on top of them, we simulate the discrete-time SIR epidemic process described in Sec.~\ref{sec.epidmod}. 

In Fig. 3a-b, we display the fraction of recovered people $R$ at the final stage in the $\beta^I-\beta^E$ plane, obtained from our numerical simulations for homogeneous (ER) and heterogeneous (SF) external connections. From these figures, we can observe two different phases: an epidemic phase in which a macroscopic fraction of the population became infected ($R>0$), and a disease-free phase, in which only a negligible number of people were infected ($R\approx 0$). The boundary separating these phases defines the critical curve $(\beta^I_c,\beta^E_c)$. 

Deriving this critical curve exactly using the generating function technique would be useful because it would allow us to quantify how the number of workers $w$ and external connectivity affect the epidemic threshold. However, finding a general analytical expression for $\beta_c^E$ that works for any value of $\beta^I$ is challenging.  For this reason, we focus on two limiting cases in which the problem simplifies considerably and analytical results can be obtained, namely $\beta^I=1$ and $\beta^I\approx 0$. The first corresponds to maximal within-household transmission, while the second describes the regime of very weak internal transmission. Although the limit $\beta^I\approx 0$ may not represent a fully realistic scenario, the expression obtained in this case can be extrapolated to small but non-negligible values of $\beta^I$, where it still provides a reasonable approximation to the epidemic threshold. Both cases are discussed in detail below.

\begin{figure}[htbp]
\centering
\panelgraphics{0.48}{(a)}{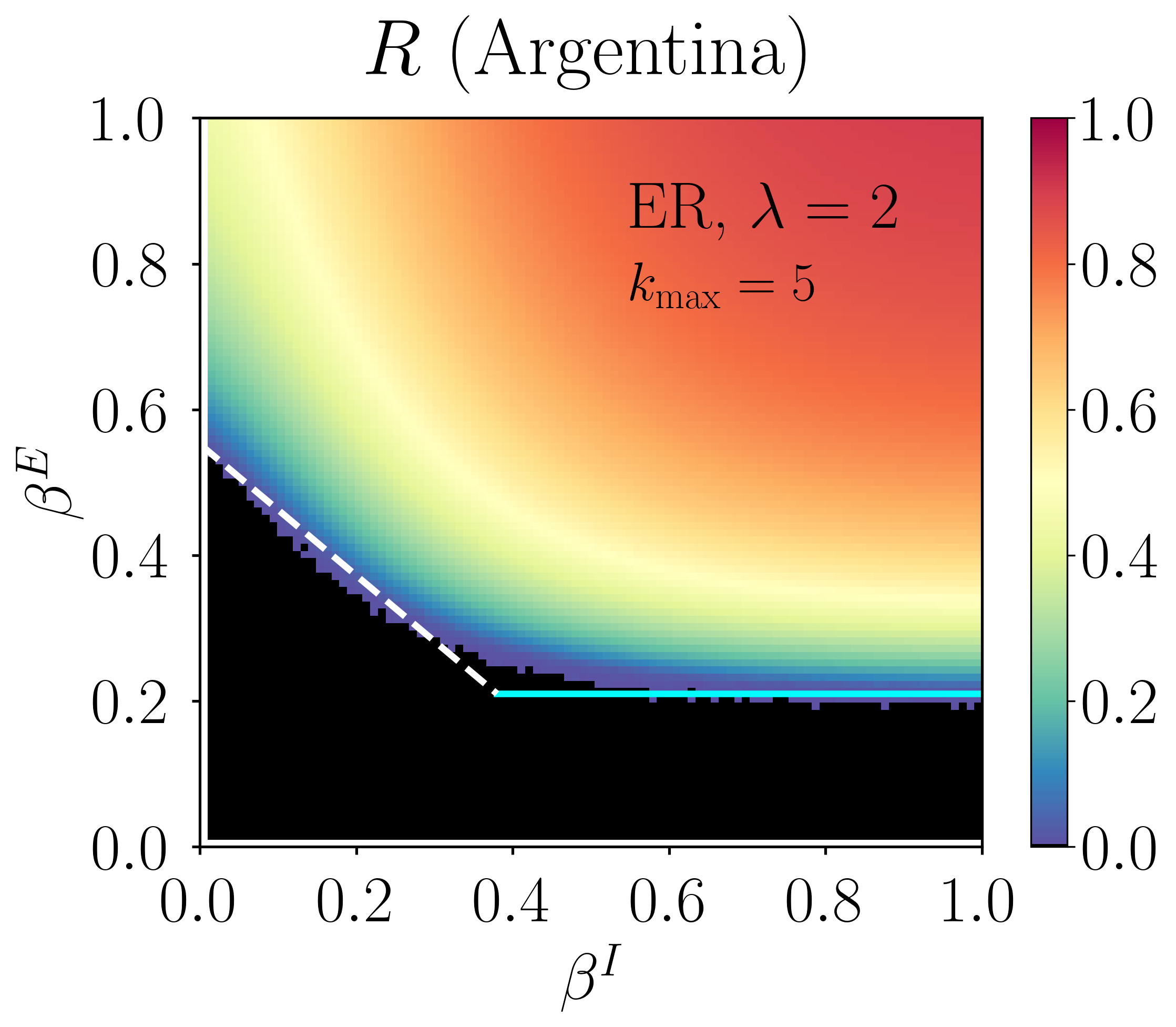}
{Heatmap of the final fraction of recovered people in the beta I beta E plane for Poisson external connections.}
\hfill
\panelgraphics{0.48}{(b)}{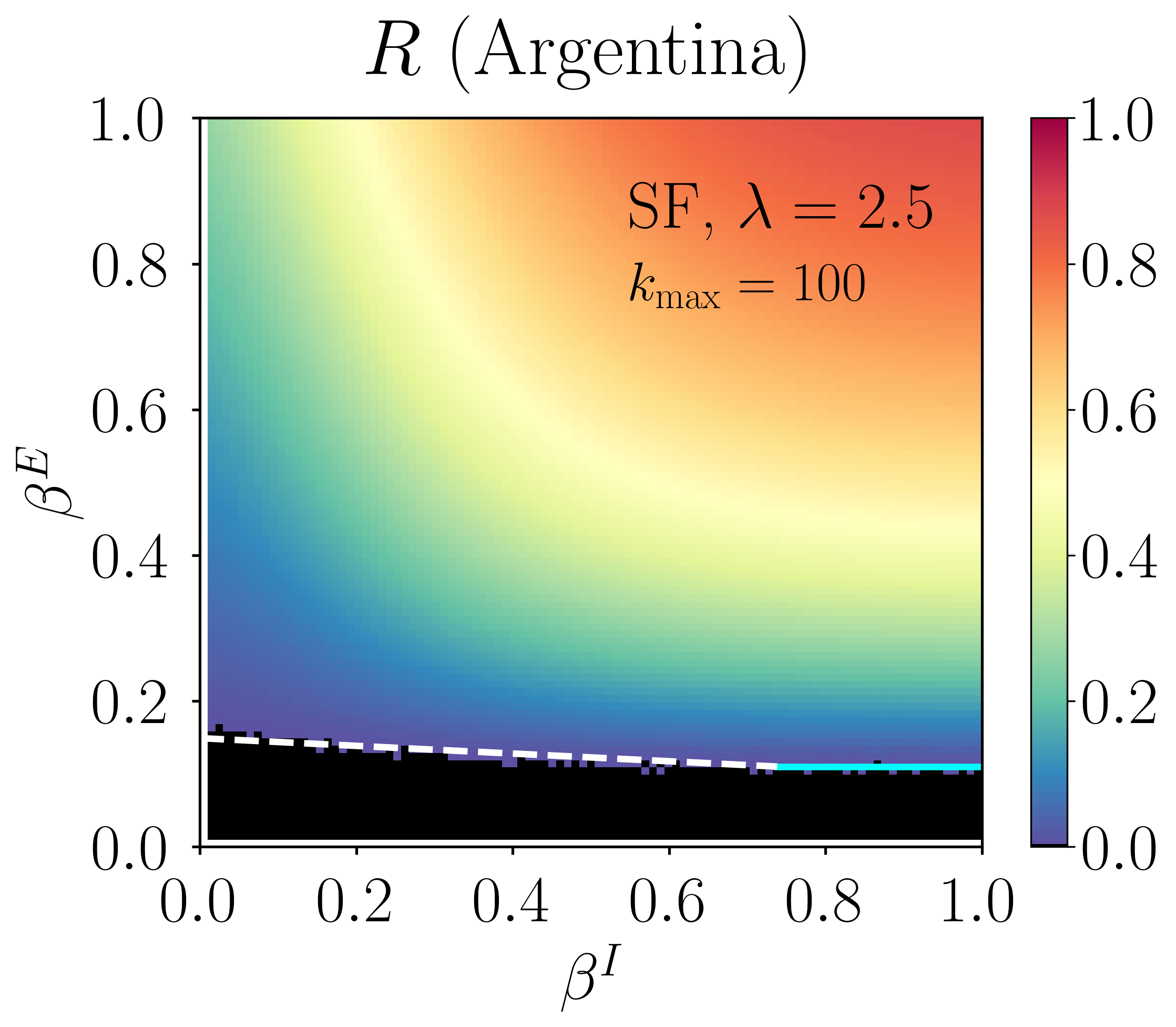}
{Heatmap of the final fraction of recovered people in the beta I beta E plane for power-law external connections.}
\caption{Fraction of recovered people $R$ at the final stage in the $\beta^I-\beta^E$ plane for Poisson (panel a) and power-law (panel b) external connections. Simulation results were averaged over 1000 stochastic realizations on networks with $N=5\times 10^5$ individuals. To compute $R$, we exclude those realizations without epidemic outbreaks ($R < 0.5$\%.). The dashed and solid lines show the analytical predictions of Eqs.~(\ref{eq.betaEcbetaIapprox0}) ($\beta^I\approx 0$) and (\ref{eqBetaEc}) ($\beta^I=1$), respectively, each extrapolated beyond its regime of validity up to the point where the two predictions intersect ($\beta^I\approx 0.4$ in panel a and $\beta^I\approx 0.7$ in panel b).}
\label{fig.NoMerg}
\end{figure}

\subsubsection{Case $\beta^I=1$}
The limit $\beta^I=1$ corresponds to the case of maximal within-household transmission in which once a single individual gets infected in a household, then the entire household becomes infected with probability one. 

One important consequence of this limit case is that, at the end of the propagation process, all members in each household are either completely susceptible or completely recovered. As a result, households can be treated as atomic units with two possible states: fully susceptible or fully recovered, and the epidemic process can be viewed as a contagious process between households rather than individuals. In other words, the epidemic process can be reformulated at the household level, which simplifies the calculations needed to study the epidemic threshold. 

By using the generating function technique (see Appendix~\ref{app.seccliqueBetaEc} for details), we find that the critical external transmission is given by,
\begin{eqnarray}\label{eqBetaEc}
\beta_c^E=\frac{1}{\frac{\langle k_E^2\rangle-\langle k_E\rangle}{\langle k_E \rangle}+\langle k_E \rangle \frac{\langle w^2\rangle-\langle w\rangle}{\langle w\rangle}}.
\end{eqnarray}

This expression shows that the critical value $\beta^E_c$ is governed by two independent sources of heterogeneity. The first term in the denominator depends solely on the heterogeneity of workers' external connectivity, measured by $\langle k_E^2\rangle-\langle k_E\rangle$, in agreement with classical results showing that epidemic thresholds are controlled by degree heterogeneity in random networks~\cite{pastor2015epidemic}. The second term depends on the variability in the number of workers per household, quantified by $\langle w^2\rangle-\langle w\rangle$, and weighted by the factor $\langle k_E\rangle$. This contribution represents the enhanced infection and transmission potential associated with households containing a larger number of workers.

Although both sources of heterogeneity lower the epidemic threshold $\beta_c^E$, they are not expected to contribute equally in real networks. Because the household size distribution $P(s)$ decays exponentially (as shown in Fig.~\ref{fig.PsPwsArg}), this imposes a limit on the variability in household size and consequently in the number of workers per household. We therefore expect the quantity $\langle w^2\rangle-\langle w\rangle$ to remain relatively small in real household networks. As a result, a lower epidemic threshold $\beta_c^E$ will be mainly driven by highly connected workers (i.e., high values of $\langle k_E^2\rangle - \langle k_E\rangle$) rather than by the variability in the number of workers within households (i.e., $\langle w^2\rangle-\langle w\rangle$). In the limit where $\langle k_E^2\rangle - \langle k_E\rangle$ is much larger than $\langle w^2\rangle-\langle w\rangle$, Eq.~(\ref{eqBetaEc}) reduces to
\begin{eqnarray}\label{eqBetaEclimit}
\beta_c^E=\frac{1}{\frac{\langle k_E^2\rangle}{\langle k_E \rangle}-1}.
\end{eqnarray}

\subsubsection{Case $\beta^I\approx 0$}

At the opposite limit, when $\beta^I$ is close to zero, within-household transmission is
very weak, and in this regime, infected workers can transmit the disease to at most one
other household member.  As a consequence, multiple within-household infections and
their probabilities do not need to be considered, which again simplifies the analysis.  Applying the generating-function technique in this regime (see Appendix~\ref{app.sec.betIapprox0}) gives
\begin{eqnarray}\label{eq.betaEcbetaIapprox0}
\beta^E_c=\frac{\langle k_E \rangle}{\langle k_E^2\rangle-\langle k_E \rangle}-\langle k_E\rangle \left(\frac{\langle w^2\rangle-\langle w\rangle}{\langle w\rangle}\right)\left(\frac{\langle k_E\rangle}{\langle k_E^2 \rangle-\langle k_E \rangle}\right)^2\beta^I. 
\end{eqnarray}
The two terms have a simple interpretation. The first is the epidemic threshold the network would have if households played no role, and the second is a correction that grows linearly with $\beta^I$ and lowers the threshold. The negative sign is the one we expect, since even the rare infections that leak inside a household add transmission pathways and make an outbreak slightly easier to trigger.

The size of this correction, equivalently the slope of the critical line in $\beta^I$, is controlled by the external degree distribution through the single factor
\begin{eqnarray}\label{eq.externalSlopeFactor}
\chi_E \equiv
\langle k_E\rangle\left(\frac{\langle k_E\rangle}{\langle k_E^2 \rangle-\langle k_E \rangle}\right)^2 = \langle k_E\rangle\,\kappa^{-2},
\end{eqnarray}
where $\kappa=\langle k_E^2 \rangle/\langle k_E \rangle-1$ is the branching factor~\cite{lopez2007limited}. Because the external heterogeneity $\langle k_E^2 \rangle-\langle k_E \rangle$ appears squared in the denominator, $\chi_E$ shrinks quadratically as the connectivity becomes more heterogeneous, and the dependence on $\beta^I$ is suppressed far faster than the threshold itself. In the limit $\langle k_E^2 \rangle-\langle k_E \rangle\to\infty$ the correction vanishes and Eq.~(\ref{eq.betaEcbetaIapprox0}) reduces to
\begin{eqnarray}\label{eq.betaEcbetaIapprox0limit}
\beta^E_c=\frac{1}{\frac{\langle k_E^2\rangle}{\langle k_E \rangle}-1},
\end{eqnarray}
the same value found in the opposite limit $\beta^I=1$, Eq.~(\ref{eqBetaEclimit}). In other words, when a few highly connected workers dominate the spreading, the epidemic threshold becomes essentially insensitive to what happens inside households.

This picture is confirmed by the simulations in Figs.~\ref{fig.NoMerg}a and b. In both networks the measured critical points vary approximately linearly with $\beta^I$ near $\beta^I\approx 0$, but with markedly different slopes, as predicted by $\chi_E$. We find $\chi_E\simeq 0.68$ for the ER network and $\chi_E\simeq 0.04$ for the SF network, so the ER boundary (panel a) is clearly tilted while the SF boundary (panel b) is almost flat. Remarkably, the linear behavior predicted by Eq.~(\ref{eq.betaEcbetaIapprox0}) (white dashed line) keeps describing this boundary accurately well beyond the regime $\beta^I\approx 0$ for which it was derived.

On the other hand, at higher values of $\beta^I$, the boundary between the epidemic and the epidemic-free phases is instead described by the constant threshold obtained in the $\beta^I=1$ limit, Eq.~(\ref{eqBetaEc}) (light-blue solid line). This value is strictly exact only at $\beta^I=1$, but the simulated boundary remains close to it as $\beta^I$ decreases, down to $\beta^I\approx 0.4$ for the ER network and $\beta^I\approx 0.7$ for the SF network, where it meets the linear prediction.

In the following section, we will explore how different merging household strategies will affect the epidemic threshold. It is important to note that, although Eqs.~(\ref{eqBetaEc}) and (\ref{eq.betaEcbetaIapprox0}) were derived using the original distribution $P(w,s)$, the same expressions apply when households are allowed to merge. In that case, all moments appearing in these equations (such as $\langle w \rangle$ and $\langle w^2 \rangle$) are computed using the merged distribution $\widetilde{P}(w,s)$ instead of $P(w,s)$. 

\subsection{Results for the bubble merging strategy}\label{sec.ResulMerg}

We now turn our attention to the scenario where cliques are merged according to the strategies described in Sec.~\ref{sec.bubstratt}. We focus on the case in
which merging is performed before the disease starts spreading, on networks
with  $P(k_E)=Pois(2,1,5)$. Results for SF networks and for other ER
parameterizations are provided in the Supplementary Material.

To quantify how many individuals benefit from each merging strategy, we denote by $f_{\mathrm{grow}}$ the fraction of individuals whose bubble size increases after merging, and the explicit calculation of this quantity for each merging strategy is given in Appendix~\ref{app.calcPtilde}.

\begin{figure}[htbp]
\centering
\panelgraphics{0.32}{(a)}{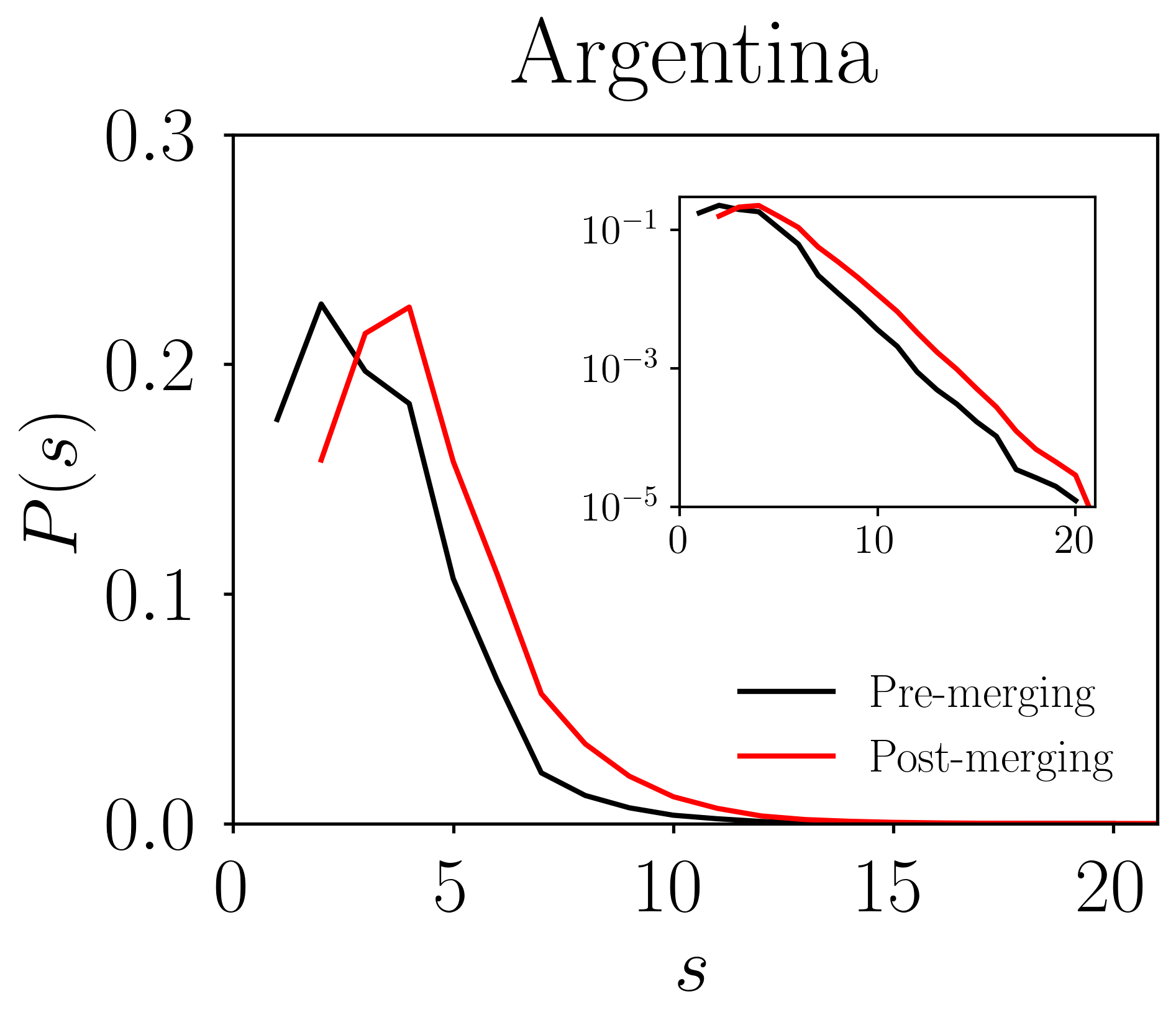}
{Household size distribution before and after the merging strategy w star equals one.}
\hfill
\panelgraphics{0.32}{(b)}{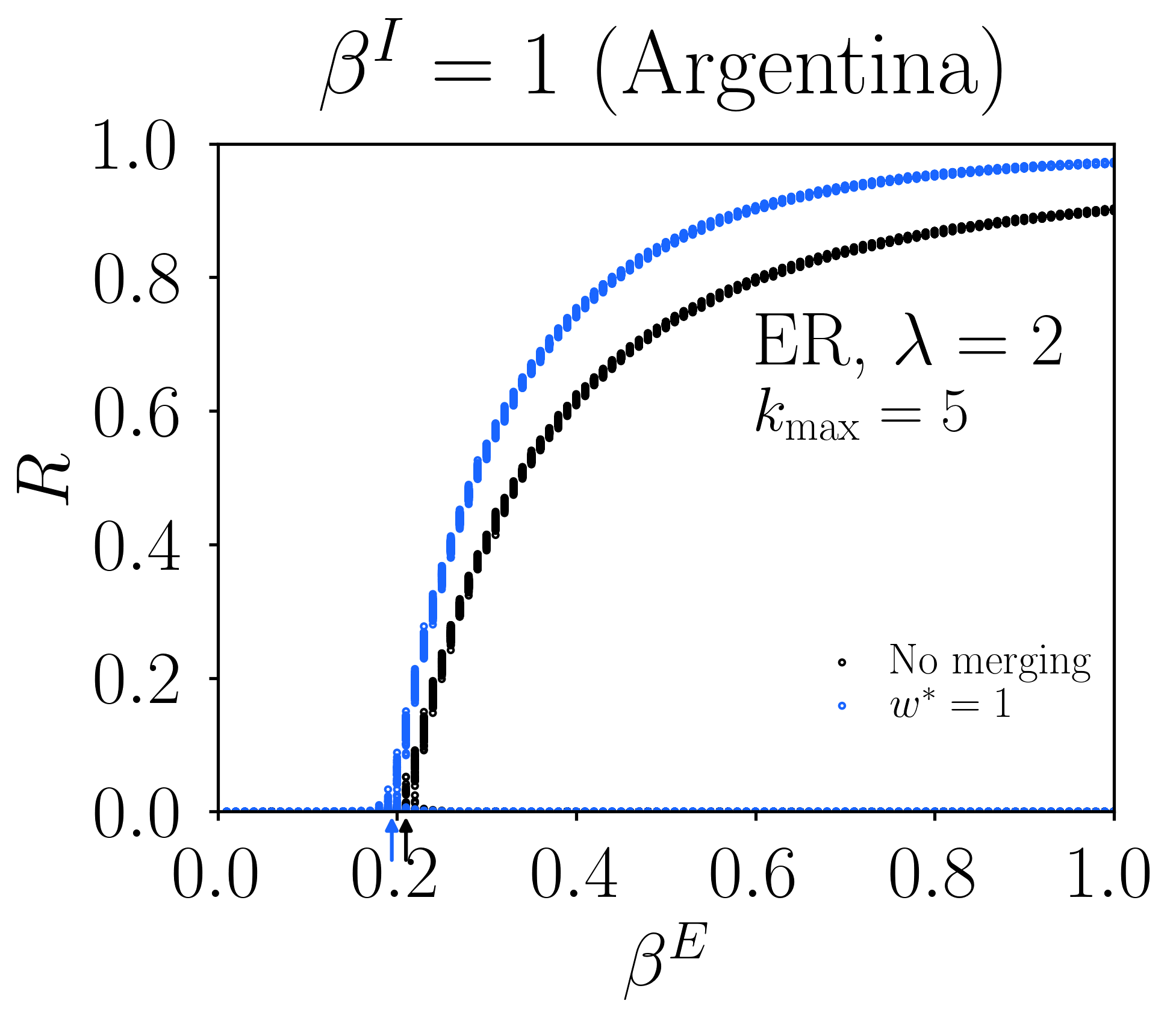}
{Final fraction of recovered people as a function of beta E for beta I equal to one.}
\hfill
\panelgraphics{0.32}{(c)}{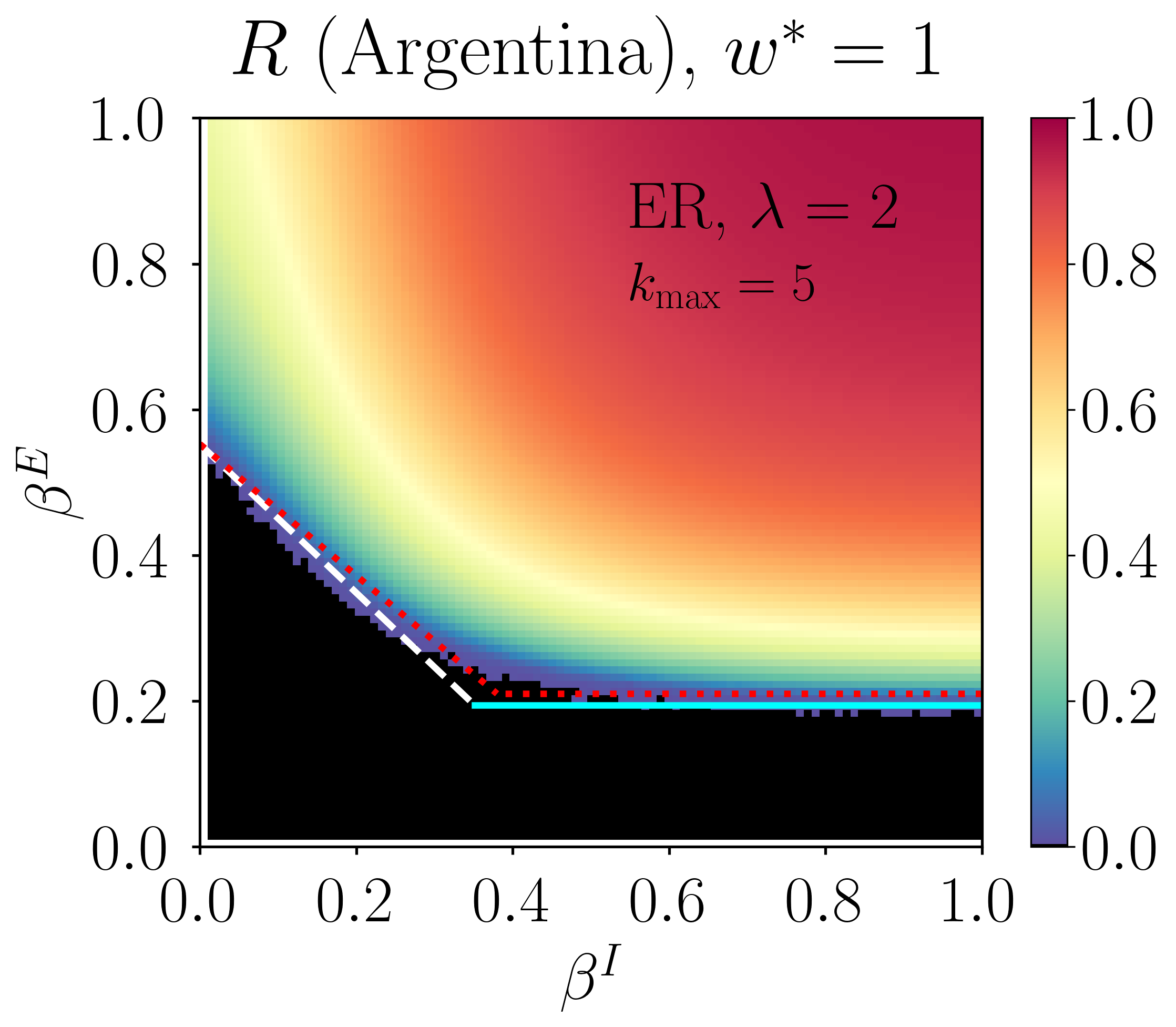}
{Heatmap of the final fraction of recovered people after the merging strategy w star equals one.}
\caption{Results for the household merging strategy $w^*=1$ in Argentina. Panel a: household size distribution $P(s)$ before merging (black) and after merging (red). The inset shows the same distributions in log-linear scales. Panel b: Scatter-plot of the fraction of recovered people $R$ at the final stage as a function of $\beta^E$ for $\beta^I=1$. Black circles correspond to the case without household merging, while blue circles correspond to the merging scenario $w^*=1$. Panel c: Heatmap of the final fraction of recovered people in the $\beta^I-\beta^E$ plane, restricted to realizations in which epidemics occur. The white dashed line shows the analytical approximation of the critical curve for $\beta^I\approx 0$ and for the household merging strategy $w^*=1$. This line was obtained from Eq.~(\ref{eq.betaEcbetaIapprox0}). Similarly, the solid line indicates the critical value $\beta^E_c$ for $\beta^I\approx 1$ after the merging strategy is applied. For comparison, the red dotted lines correspond to the critical curves without household merging shown in Fig.~\ref{fig.NoMerg}a.}
\label{fig.fusionN1}
\end{figure}

We begin with Scenario $w^*=1$, in which households with at most one worker are merged in pairs. Figure~\ref{fig.fusionN1}a presents the distribution of household sizes before and after applying the merging strategy, illustrating, as expected, a clear shift toward larger household sizes. Because these newly formed bubbles contain more workers, and therefore possess a higher number of external connections, it is natural to expect an increase in epidemic risk (i.e., a lower value of $\beta^E_c$). The relevant question, however, is whether this increase remains moderate or not.

To answer this question, we compared the cases with and without household merging, and used the theoretical equations that predict the critical epidemic threshold given in the previous section. In Fig.~\ref{fig.fusionN1}b, we show the fraction of recovered people $R$ at the final stage for $\beta^I=1$, and assuming external connections that follow a truncated Poisson distribution $Pois(2,1,5)$. As expected, from this figure, we can see that merging households leads to a larger fraction of infected people. Nevertheless, by using Eq.~(\ref{eqBetaEc}), we find that the critical external transmission probability for the $w^*=1$ scenario is $\beta_c^E=0.194$, which is slightly lower than the value $\beta^E_c=0.21$ obtained for the case without merging.

This small change in the value of $\beta_c^E$ is not just limited to the case $\beta^I=1$. As shown in  Fig.~\ref{fig.fusionN1}c, the shift of the critical curve remains moderate across the entire range of $\beta^I\in [0,1]$, suggesting that the Scenario $w^*=1$ does not dramatically alter the epidemic threshold. Therefore, from the perspective of epidemic control, this result indicates that the relative increase in epidemic risk could plausibly be mitigated by complementary non-pharmaceutical measures, like contact tracing and social distancing.

Beyond disease control, this strategy also has a clear effect on social connectivity. For the $w^*=1$ scenario, the average number of internal links rises from $\langle \ell_I \rangle = 5.4$ (no merging) to  $\langle \ell_I \rangle = 9.7$  (see Table~\ref{tab.Resumen}). In addition, approximately $f_{\mathrm{grow}}\simeq 40\%$ of individuals end up in a larger bubble.

\begin{figure}[htbp]
\centering
\panelgraphics{0.48}{(a)}{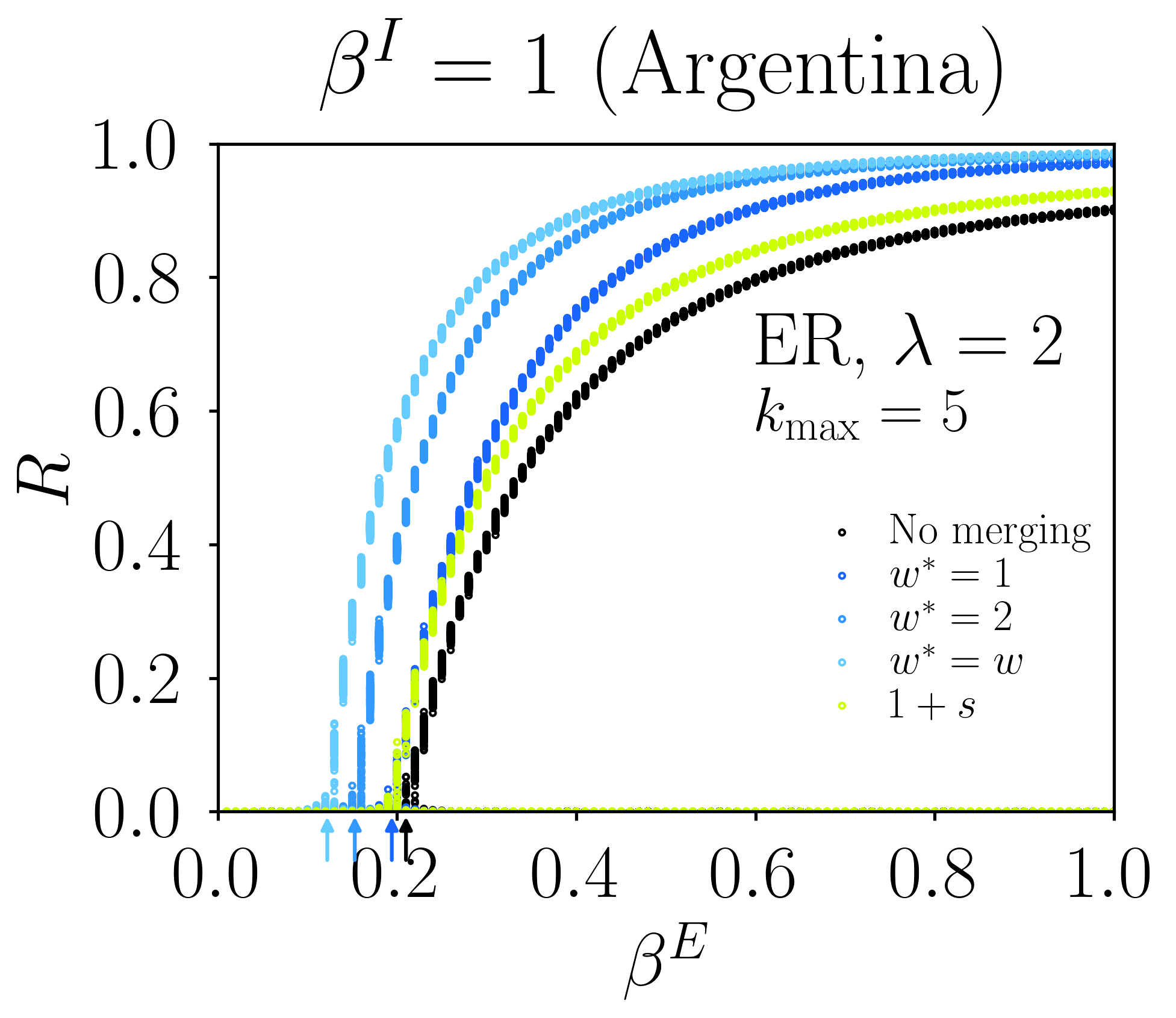}
{Final fraction of recovered individuals as a function of beta E for all merging strategies.}
\hfill
\panelgraphics{0.48}{(b)}{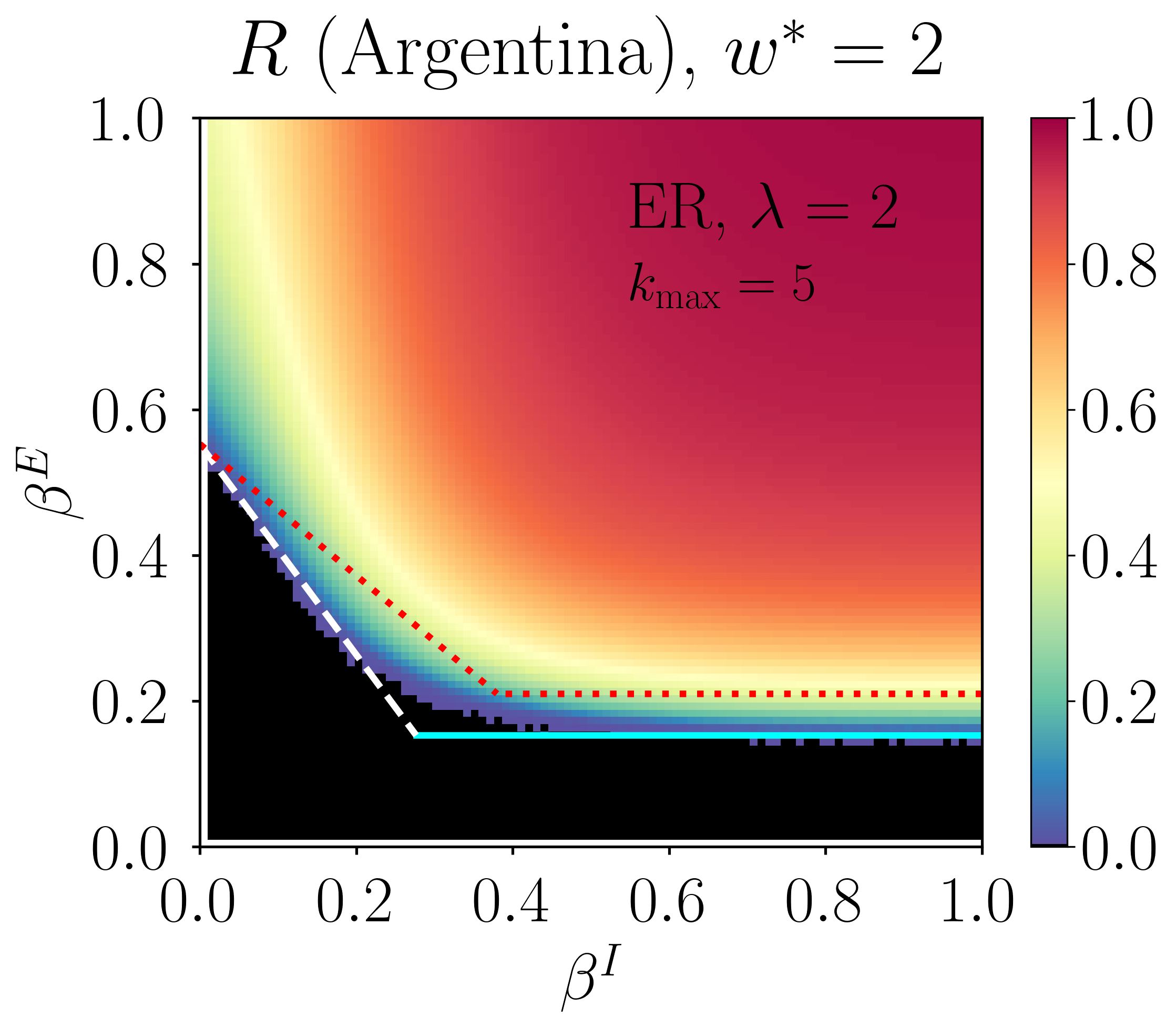}
{Heatmap for the merging strategy w star equals two.}

\vspace{1ex}

\panelgraphics{0.48}{(c)}{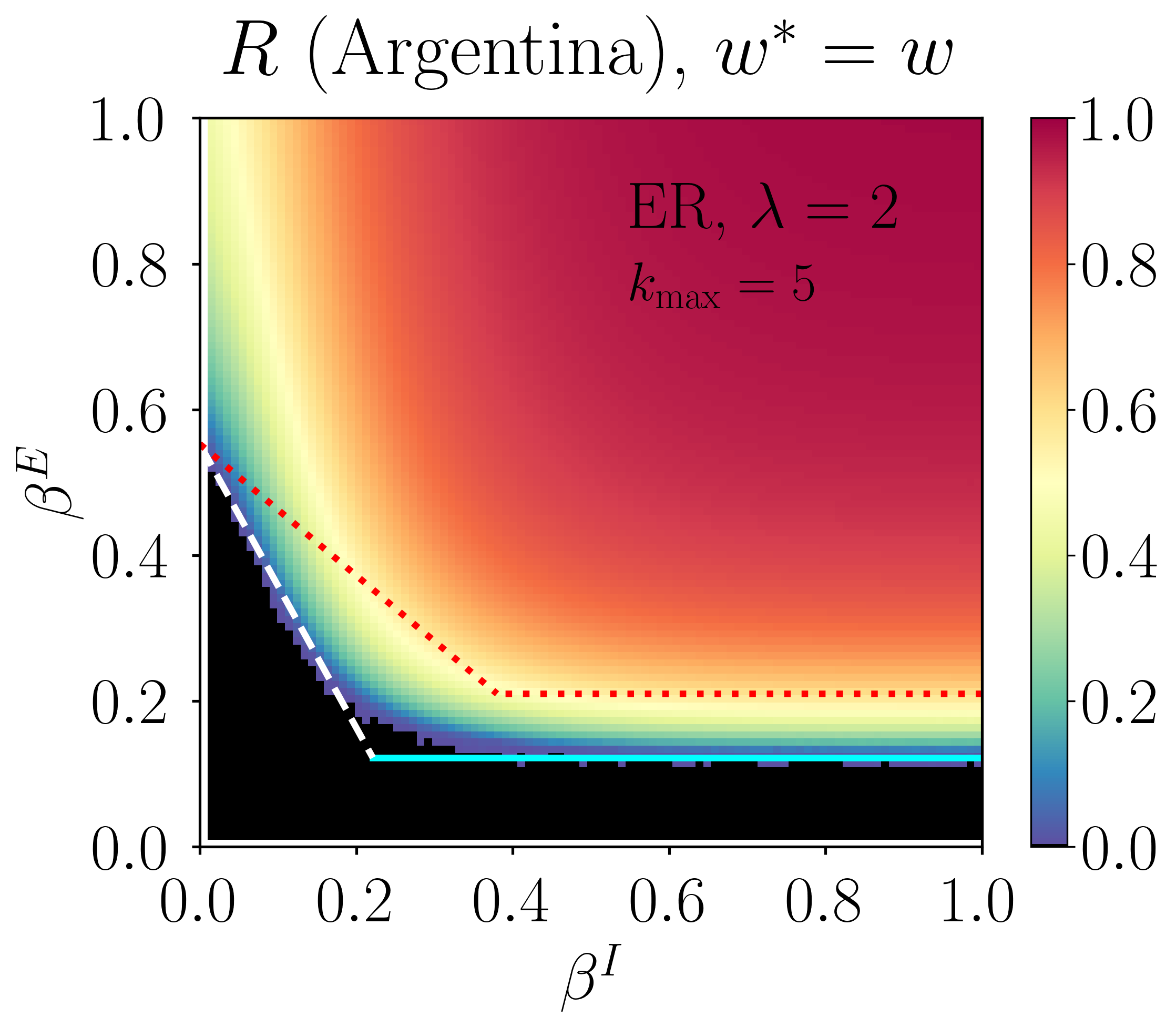}
{Heatmap for the merging strategy where all households merge regardless of the number of workers.}
\hfill
\panelgraphics{0.48}{(d)}{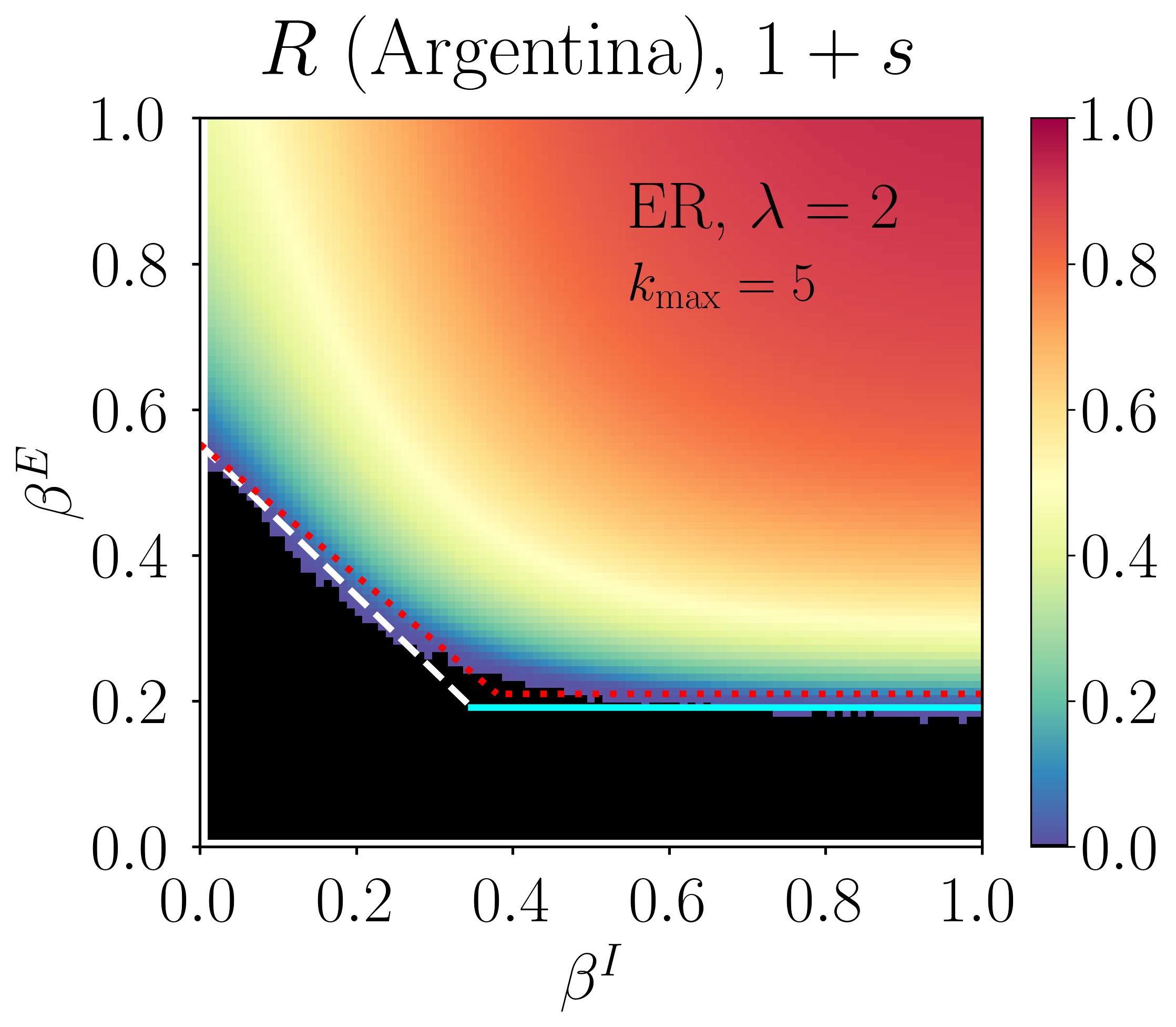}
{Heatmap for the household-size-based merging strategy one plus s.}
\caption{Results for the remaining household merging strategies. Panel a: scatter-plot of the fraction of recovered individuals $R$ at the final stage as a function of $\beta^E$ (with $\beta^I=1$) for all merging strategies, including the case without merging. Panels b-d: Heatmaps of the final fraction of recovered individuals in the $\beta^I-\beta^E$ plane, for the following merging strategies: $w^*=2$ (panel b), $w^*=w$ (panel c), and $1+s$ (panel d). The white dashed lines show the analytical approximation of the critical curve for $\beta^I\approx 0$. These lines were obtained from Eq.~(\ref{eq.betaEcbetaIapprox0}). Similarly, the solid lines indicate the critical value $\beta^E_c$ for $\beta^I\approx 1$ after the merging strategy is applied. For comparison, the red dotted lines correspond to the critical curves without household merging shown in Fig.~\ref{fig.NoMerg}a.}
\label{fig.fusionNOthers}
\end{figure}

We will next explore the remaining merging scenarios (see Figures~\ref{fig.fusionNOthers}a-d and Table~\ref{tab.Resumen}). On one hand, we obtain that under Scenario $w^*=2$, the increase of the number of internal connections is significantly larger, reaching $\langle \ell_I\rangle=15.8$, and at least 77\% of the population ends up in a larger bubble. These effects are even more pronounced under Scenario $w^*=w$. These two scenarios are therefore likely to have a stronger positive psychological effect than Scenario $w^*=1$. However, they also lead to a substantial reduction in the epidemic threshold $\beta_c^E$, as shown in Fig.~\ref{fig.fusionNOthers}a and Table~\ref{tab.Resumen}. 

Finally, we consider two additional merging scenarios based on household size. In the $1+s$ scenario, every single-person household merges with another randomly chosen household of size two or larger (see Sec.~\ref{sec.bubstratt}). For $\beta_I=1$, the critical external transmission probability is  $\beta^E_c=0.189$, which is nearly the same as in Scenario $w^*=1$ (see Figure~\ref{fig.fusionNOthers}a and Table~\ref{tab.Resumen}). As a consequence, from an epidemic control point of view, both strategies lead to a similar level of epidemic risk. However, the social effects are different. In Scenario  $1+s$, the increase in internal connectivity is smaller, with $\langle \ell_I \rangle = 7.4$ (Table~\ref{tab.Resumen}), and only about 26\% of individuals experience an increase in bubble size. 

We also consider Scenario $2+s$, in which every household with at most two people merges with another randomly chosen household of size three or larger. This strategy achieves $\langle \ell_I\rangle=13.7$ and benefits approximately 74\% of the population with larger social bubbles (Table~\ref{tab.Resumen}). However, it also leads to $\beta^E_c=0.146$, which represents a more substantial reduction in the epidemic threshold compared to Scenarios $1+s$ and $w^*=1$.

Taken together, these results indicate that the $w^*=1$ scenario provides the most balanced outcome among the strategies considered because it allows a substantial portion of the population to join in larger social bubbles without dramatically increasing epidemic risk.

\begin{table}[htbp]
  \centering
  \caption{Summary of the main results for the different household merging strategies in Argentina. All reported values correspond to the case of external connections following a Poisson degree distribution $Pois(2,1,5)$. For each scenario, we report the critical external transmission probability $\beta^E_c$ (for $\beta^I=1)$, the average number of internal links $\langle \ell_I\rangle$, and the fraction of the population whose bubble size increases after merging, $f_{\mathrm{grow}}$.  The calculation of $f_{\mathrm{grow}}$ is given in Appendix~\ref{app.calcPtilde}. The case without merging is included for comparison.}
    \begin{tabular}{l@{\hspace{25pt}}c@{\hspace{25pt}}c@{\hspace{25pt}}c}
        \hline
          Scenario        & $\beta^E_c$   & $\langle \ell_I\rangle$  & $f_{\mathrm{grow}}$        \\
        \hline
          No merging      & 0.21         & 5.4          & ---   \\
          $w^*=1$         & 0.194        & 9.7          & 41\%  \\
          $w^*=2$         & 0.152        & 15.8        & 77\%   \\
          $w^*=w$         & 0.122        & 21.4           & 100\%  \\
          $1+s$           & 0.189        & 7.4          & 26\%  \\
          $2+s$           & 0.146        & 13.7          & 74\%  \\
        \hline
    \end{tabular}
  \label{tab.Resumen}
\end{table}

\section{Conclusion}
In this work, we have studied different household bubbling strategies and their effects on both epidemic control and social connectivity. While previous studies have proposed merging household strategies based on a variety of criteria, including household size, age composition, or random pairing, here we focus on the number of economically active members in each household. This information is usually available in census micro-data but, to the best of our knowledge, has not been used in the context of social bubble design.

We first analyzed a baseline scenario in which households do not merge. In this setting, we explored when an epidemic can or cannot spread, focusing especially on two extreme cases: when infections inside households are very unlikely ($\beta^I \approx 0$) and when they spread very easily within households ($\beta^I=1$). For both cases, we used the generating function technique to estimate the epidemic threshold, obtaining results that agree well with our numerical simulations. Notably, even though these are limiting cases, we also found that they approximate very well the epidemic threshold curve across the whole range of internal transmission probabilities ($\beta^I\in [0,1]$). 

Then, we explored several merging approaches based on the number of workers in each household, and compared them with two merging strategies based on household size. Our results showed that merging households with at most one worker provides the best balance between epidemic control and social connectivity. This strategy has a similar epidemic risk to the $1+s$ strategy, but it allows more people to be part of larger social bubbles. We observed the same qualitative results across all countries considered in this work, which suggests that this strategy performs well in different demographic contexts.

As with all modeling approaches, this work has several limitations, and our results should be interpreted with caution. First, we assumed that each worker has a fixed and randomly distributed number of external contacts. A more realistic model would explicitly include structured workplace groups, but this would require detailed data that is often difficult to obtain. On the other hand, in our model, we do not distinguish between essential and non-essential workers, and instead allow all workers to have contacts outside their households. This scenario is different from what happened during the COVID-19 lockdown, where only essential workers were allowed to go to work. Finally, our main results rely on the assumption that non-working household members remain isolated. However, if non-workers also form external contacts, the epidemic risk of worker-based merging strategies could rise substantially. In that case, merging strategies based on household sizes or a combination of different merging strategies may offer better control of disease spread.  

Despite these assumptions and limitations, our work shows that already available data can be used to explore alternative and more flexible bubbling strategies.  We hope that this work will contribute to the development of more socially sustainable non-pharmaceutical interventions for future epidemics.

\section{Acknowledgments}\label{Sec.Ack}
This work was partially funded by CONICET. The authors wish to acknowledge the following statistical offices that provided the underlying data making this research possible: National Institute of Statistics and Censuses (Argentina), National Institute of Statistics (Spain), Central Bureau of Statistics (Israel), and National Bureau of Statistics (China). We also thank the anonymous reviewers for their valuable comments and suggestions to improve this paper.

\appendix

\section{Derivation of the epidemic critical point $\beta^E$ for the SIR model on networks with bubbles}
In this appendix, we will derive the critical point $\beta_c^E$ for our epidemic model on networks with bubbles in two limiting cases: $\beta^I=1$ (perfect intra-bubble transmission) and $\beta^I\approx 0$ (almost negligible intra-bubble transmission). To provide the necessary background for our derivation, we will first briefly review how the generating function technique is used to study epidemics in networks without cliques, and then we will apply this technique to estimate $\beta_c^E$ in our model.

\subsection{Background: the SIR model on networks without cliques}\label{sec.appSIRclassic}

One of the most common models for studying diseases that confer permanent immunity is the SIR model. In its discrete-time version, this model involves two transitions:
\begin{itemize}
\item $I+S \xrightarrow{\beta} I+I$, meaning that a susceptible person who is in contact with an infected neighbor can become infected at the next time step with probability $\beta$.
\item $I \xrightarrow{t_r} R$, meaning that an infected person moves deterministically to the recovered compartment after $t_r$ time steps.
\end{itemize}
Typically, the process begins with a single infected individual, also called the {\it index-case}, and then, from this initial condition, the disease starts spreading through the contact network until no infected people remain~\cite{spricer2019sir}. At this final stage, each person can be in only one of the two following states: susceptible or recovered. In the language of Statistical Mechanics, the final fraction of recovered individuals $R$ is the order parameter of this model, and $\beta$ serves as the control parameter (assuming $t_r=1$)~\cite{pastor2015epidemic,stauffer2018introduction}. It is well known that the SIR model undergoes a second-order phase transition at a critical threshold $\beta=\beta_c$, below which there are no epidemics and only a microscopic fraction of nodes become infected (i.e., $R\approx 0$), while above $\beta_c$, a macroscopic fraction of the population can be infected ($R>0$). In other words, we say that above $\beta_c$, large outbreaks or epidemics are possible. An important feature of the critical point is that its value depends on the network topology~\cite{pastor2015epidemic}. While the estimation of $\beta_c$ for spatial networks is usually accessible only through numerical simulations, for random networks, $\beta_c$ can be calculated exactly by using the generating function technique~\cite{pastor2015epidemic,newman2002spread}, which we will describe below.

Consider a random network (with a locally tree-like structure) where nodes have a degree or connectivity $k$ drawn from a probability distribution $P(k)$. From this probability, the following two generating functions are defined:
\begin{itemize}
\item $G_0(x)=\sum_k P(k) x^k$,
\item $G_1(x)\equiv \langle k\rangle^{-1}\frac{dG_0(x)}{dx}=\sum_k \frac{k P(k)}{\langle k \rangle} x^{k-1}$,
\end{itemize}
where $\langle k \rangle=\sum_k kP(k)$ is the average degree, $G_0(x)$ is the generating function for the probability that a node has $k$ connections, and $G_1(x)$ is the generating function for the excess-degree, which represents the number of remaining links when one arrives at a node by following a randomly chosen link.

As explained in detail in Refs.~\cite{wang2017unification,dall2005inhomogeneous}, for random networks, the final fraction of recovered people $R$ can be calculated as a function of $\beta$ by using the generating functions given above. Specifically, the fraction $R$ is obtained by solving the following two equations
\begin{eqnarray}
q_{\infty}&=&1-\beta+\beta G_1(q_{\infty}),\label{eq.appqinf}\\
R&=&1-G_0(q_{\infty}),
\end{eqnarray}
where $q_{\infty}$ represents the probability that a node (reached by following a randomly chosen link) remains susceptible or recovers without having been the source of a large outbreak. Note that $q_{\infty}=1$ is always a solution of Eq.~(\ref{eq.appqinf}) for any value of $\beta$, because $G_1(1)=1$. This value $q_{\infty}=1$ corresponds to the case where no epidemic occurs ($R=0$) and it is known as the trivial solution. For $\beta<\beta_c$ only the trivial solution exists, meaning that in this regime, any outbreak remains small and finite. However, for $\beta>\beta_c$, a second non-trivial solution emerges which corresponds to the physical solution of this SIR model, and for this case we have a macroscopic fraction of recovered individuals $R>0$. The critical point $\beta=\beta_c$ represents the boundary between these two regimes. Graphically, this is the point where the right-hand side of Eq.~(\ref{eq.appqinf}) becomes tangent to the identity line $q_{\infty}$ at the point $q_{\infty}=1$. This tangency condition allows us to calculate $\beta_c$ by differentiating both sides of Eq.~(\ref{eq.appqinf})  with respect to $q_{\infty}$ and setting $q_{\infty}=1$:
\begin{eqnarray}
1&=& \beta_c \frac{d G_1 (q_{\infty})}{dq_{\infty}}\Bigr|_{q_{\infty}=1}.
\end{eqnarray}
Since $\frac{d G_1 (q_{\infty})}{dq_{\infty}}\Bigr|_{q_{\infty}=1}=(\langle k^2\rangle - \langle k\rangle)/\langle k\rangle$, we finally obtain,
\begin{eqnarray}\label{eq.appbc}
\beta_c&=&\frac{\langle k\rangle}{\langle k^2\rangle-\langle k\rangle}.
\end{eqnarray}
This expression reveals that the epidemic threshold in random networks is governed by the first and second moments of the degree distribution. In homogeneous networks, where most nodes have similar degrees, these two quantities are of the same order, so the critical transmission probability $\beta_c$ remains finite. In contrast, highly heterogeneous networks are typically characterized by a small number of nodes (known as hubs) with very large degrees, while the vast majority of nodes have small degrees. As a result, $\langle k^2\rangle$ is much larger than $\langle k\rangle$, causing $\beta_c$ to approach zero.

\subsection{The SIR model on networks with cliques for $\beta^I=1$}\label{app.seccliqueBetaEc}

In this section, we will calculate the critical value $\beta_c^E$ for our SIR model on networks with cliques or bubbles. As explained in Sec.~\ref{sec.epidmod}, for this model, there are two transmission probabilities, namely $\beta^I$ and $\beta^E$, which are the infection probabilities within and outside cliques, respectively. In order to simplify the analytical calculations, we will focus here only on the case where the internal infection probability is $\beta^I=1$, which implies that once a member in a household gets infected, then every susceptible person within the same household will also become infected. While this represents the most extreme scenario of transmission within a confined space, it is a reasonable simplification for diseases that spread very easily in small and highly dense environments. One immediate consequence of $\beta^I=1$ is that, at the final stage, only two outcomes are possible within a bubble: either all members remain susceptible, or all members are in the recovered state. Consequently, a household can be treated as an atomic unit of the epidemic process. From this perspective, the epidemic process at the household level becomes mathematically equivalent to the SIR model on networks without cliques, with $\beta^E$ playing the same role as the parameter $\beta$ presented in the previous section. Therefore, we can directly apply the generating function technique and adapt the equations from the previous section to compute the critical value $\beta_c^E$ for our epidemic model on networks with cliques. 

To obtain $\beta_c^E$ for our model, we must first characterize the network structure at the household level. Recall that in the SIR model without cliques, the critical point $\beta_c$ is obtained from the generating function $G_0(x)$, which encodes the degree distribution $P(k)$, that is, the fraction of nodes with $k$ connections. Now, for our model with cliques and $\beta^I=1$, because households act as atomic units, we need an analogous generating function that describes the network at the household level. Specifically, we need to characterize the total external connectivity of each household, defined as the sum of external connections of all workers in the household. For example, consider a household with two workers, one having two external connections and the other three. In this case, the household has a total external degree equal to five.

In order to obtain the distribution of this total external degree, we introduce the following two auxiliary generating functions: 
\begin{itemize}
\item $G_{0}^{E}(x)=\sum_{k_E}P(k_E)x^{k_E}$, which is the generating function for the probability that a worker in a household has $k_E$ external connections.
\item $G_{0}^{C}(x)=\sum_s\sum_w P(s)P(w|s)x^w$, which is the generating function for the probability that a household/clique contains a total number $w$ of workers.
\end{itemize}
Then, by composing these two functions, we obtain the generating function for the total number of external connections per household:
\begin{eqnarray}\label{eq.appG0T}
G_{0}^T(x)\equiv G_0^C(G_0^E(x))=\sum_s\sum_{w=0}^{s} P(s)P(w|s)(G_0^E(x))^w.
\end{eqnarray}
With this generating function in hand (which plays the same role as the function $G_{0}(x)$ presented in Sec.~\ref{sec.appSIRclassic}), we can now follow the same procedure as in the previous section to derive the critical point.

Following the analogy with an SIR model without cliques, our model with cliques now satisfies the equation,
\begin{eqnarray}
q_{\infty}=1-\beta^E+\beta^E G_1^{T}(q_{\infty}), \label{eq.appqinfBub}
\end{eqnarray}
which mirrors Eq.~(\ref{eq.appqinf}). Here, $G_{1}^T(x)$ is related to the derivative of $G_{0}^T(x)$ with respect to $x$:
\begin{eqnarray}
G_1^{T}(x) \equiv \frac{1}{\langle k_E\rangle \langle w\rangle}\frac{dG_0^T(x)}{dx},
\end{eqnarray}
where $\langle k_E\rangle$ is the average number of external connections per worker, and $\langle w\rangle$ is the average number of workers per household. Note that $G_1^{T}(1)=1$.

Then, by applying the same procedure as in the previous subsection (that is, differentiating both sides of Eq.~(\ref{eq.appqinfBub}) with respect to $q_{\infty}$ and setting $q_{\infty}=1$) we obtain, after some algebra, the critical value of the external infection probability:
\begin{eqnarray}\label{eq.appbetaE}
\beta_c^E=\frac{1}{\frac{\langle k_E^2\rangle - \langle k_E\rangle}{\langle k_E \rangle}+\langle k_E \rangle \frac{\langle w^2\rangle-\langle w\rangle}{\langle w\rangle}}.
\end{eqnarray}

\subsection{Critical threshold for $\beta^I\approx 0$}\label{app.sec.betIapprox0}
In Sec.~\ref{app.seccliqueBetaEc}, we presented the equations predicting the critical point $\beta_c^E$ for $\beta^I=1$. Now, we will analyze the opposite limit where the transmission within households is very weak ($\beta^I\approx 0$) and estimate the critical point for this case.

As in the previous two sections, we will again use the generating function technique to estimate $\beta_c^E$. 

Consider a network composed of cliques, and then we randomly choose a household with $w$ workers, one of whom is infected. This person may transmit the disease to $k_I^A$ other workers within the household. When $\beta^I$ is very small, the most probable events of transmission within this household are the following:
\begin{itemize} 
\item the infected worker does not infect any other worker in the same household ($k_I^A=0$), which occurs with probability $P(k_I^A=0|w)=(1-\beta^I)^{w-1}$,
\item the infected worker transmits the infection to at most one additional worker ($k_I^A=1$). Because the probability of infecting two or more workers within a household is negligible for $\beta^I\approx 0$, we can then approximate $P(k_I^A=1|w)=1-P(k_I^A=0|w)=1-(1-\beta^I)^{w-1}$.
\end{itemize}

Note that the probabilities given above, $P(k_I^A=0|w)$ and $P(k_I^A=1|w)$, are conditioned by the fact that we choose a household with $w$  workers. Now, given that $w P(w)/\langle w\rangle$ is the total probability that a randomly chosen worker belongs to a household with $w$ workers, we then have that the total probability $P(k_I^A=0)$ is:
\begin{eqnarray}
P(k_I^A=0)&=&\sum_{w} \frac{wP(w)}{\langle w\rangle}P(k_I^A=0|w)=\sum_{w} \frac{wP(w)}{\langle w\rangle}(1-\beta^I)^{w-1},\nonumber\\
&\approx&\sum_w \frac{wP(w)}{\langle w\rangle}(1-(w-1)\beta^I),\nonumber\\
&\approx& 1-\frac{\langle w^2\rangle -\langle w\rangle}{\langle w\rangle}\beta^I.
\end{eqnarray}
and thus, the total probability of infecting exactly one worker $P(k_I^A=1)$ is:
\begin{eqnarray}
P(k_I^A=1)&\approx&1-P(k_I^A=0),\nonumber\\
&\approx&\frac{\langle w^2\rangle -\langle w\rangle}{\langle w\rangle}\beta^I.
\end{eqnarray}.

With this probability distribution $P(k_I^A)$, we can construct the following generating functions that contain all the information of transmission events within cliques:
\begin{eqnarray}
G^A_0(x)&=&\sum_{k_I^A=0}^{1}P(k_I^A)x^{k^A_I}=\left(1-\frac{\langle w^2\rangle -\langle w\rangle}{\langle w\rangle}\beta^I\right)+\left(\frac{\langle w^2\rangle -\langle w\rangle}{\langle w\rangle}\beta^I\right)x, \\
G^A_1(x)&=&c_A\frac{d G_0^A (x)}{dx}=1.
\end{eqnarray}
where $c_A$ is a normalization constant. 

Following the standard generating function approach, we write the following self-consistent equations:
\begin{eqnarray}
q_{\infty}^E&=&1-\beta^E+\beta^E G_{1}^{E}(q_{\infty}^E)G^A_0(q_{\infty}^I),\label{eq.app.QinfE}\\
q_{\infty}^I&=&G^A_1(q_{\infty}^I)G_{0}^{E}(q_{\infty}^E),\label{eq.app.QinfI}
\end{eqnarray}
where $q_{\infty}^E$ and $q_{\infty}^I$ denote the probabilities that an external and internal link, respectively, do not lead to a large epidemic outbreak. Note that in order to obtain Eq.~(\ref{eq.app.QinfI}), we have assumed that internal connections are distributed randomly among all workers in the network. This is clearly a crude simplification, because in our model, internal links are not actually distributed at random but rather form cliques. However, because $\beta^I\approx 0$ and at most one internal transmission occurs within a clique, the infection does not "see" the full clique structure, making this approximation reasonable.

Now, substituting $q_{\infty}^I$ from Eq.~(\ref{eq.app.QinfI}) into Eq.~(\ref{eq.app.QinfE}), we obtain a self-consistent equation for $q_{\infty}^E$:
\begin{eqnarray}\label{eq.app.qEfinal}
q_{\infty}^E=1-\beta^E+\beta^E G_{1}^{E}(q_{\infty}^E)\left(1-\frac{\langle w^2\rangle-\langle w\rangle}{\langle w\rangle}\beta^I+\frac{\langle w^2\rangle-\langle w\rangle}{\langle w\rangle} \beta^I G_{1}^{E}(q_{\infty}^E)\right),
\end{eqnarray}
and then, following the same procedure as in previous sections, we can calculate the critical point $\beta^E=\beta_c^E$ by differentiating both sides of Eq.~(\ref{eq.app.qEfinal}) with respect to $q_{\infty}^E$ at the point $q_{\infty}^E=1$. Finally, after algebraic manipulations, we obtain:
\begin{eqnarray}
\beta^E_c=\frac{\langle k_E \rangle}{\langle k_E^2\rangle-\langle k_E \rangle}- \langle k_E\rangle \left(\frac{\langle w^2\rangle-\langle w\rangle}{\langle w\rangle}\right)\left(\frac{\langle k_E\rangle}{\langle k_E^2 \rangle-\langle k_E \rangle}\right)^2 \beta_I.
\end{eqnarray}

\section{Calculation of $\widetilde{P}(w,s)$ and $f_\mathrm{grow}$}\label{app.calcPtilde}

In this section, we will calculate two quantities for each merging scenario presented in Sec.~\ref{sec.bubstratt}. First, we compute how the distribution of household sizes and number of workers changes after the merging process. Recall that $P(w,s)$ denotes the fraction of households with size $s$ and $w$ workers before merging, while $\widetilde{P}(w,s)$ denotes the corresponding distribution after households merge into larger bubbles. Second, we compute the fraction of individuals $f_\mathrm{grow}$ whose bubble size increases after applying each strategy. 

\subsection{Merging strategies based on the number of working members}
In this scenario, only households with at most $w^*$ workers are allowed to merge with each other. Let $q_w$ be the fraction of households with $w$ workers that will merge with another household. For this scenario, we have $q_0=q_1=\cdots=q_{w^*}=1$ and $q_{w}=0$ for $w>w^*$. The resulting distribution after merging is,
\begin{eqnarray}
\widetilde{P}(w,s)=\frac{1}{c}\left((1-q_w)P(w,s)+\frac{1}{2} \sum_{s_1=1}^{s}\sum_{w_1=0}^{w^*}q_{w_1}P(w_1,s_1)\frac{q_{w-w_1} P(w-w_1,s-s_1)}{\sum_{w'_1=0}^{w^*}q_{w'_1}P(w'_1,s_1)}\right),
\end{eqnarray}
where:
\begin{itemize}
\item $c$ is a normalization factor to ensure that $\sum_{s=1}^{\infty}\sum_{w=0}^{s}\widetilde{P}(w,s)=1$.
\item the first term on the r.h.s. corresponds to households of size $s$ with $w$ workers that did not merge,
\item the second term on the r.h.s. is the fraction of merged bubbles that come from two households: one of size $s_1\leq s$ with $w_1\leq w$ workers, and the other household with size $s-s_1$ and $w-w_1$ workers. After these two households merge, we obtain a larger bubble with size $s$ and $w$ workers, increasing the value of the probability $\widetilde{P}(w,s)$. The factor of 2 in this second term avoids double counting of household pairs in the summation [for example, a merge between households of sizes $s_1=1$ and $s_2=2$ is counted both as ($s_1=1,s-s_1=2$) and as ($s_1=2,s-s_1=1$)]. On the other hand, the factor $\sum_{w'_1=0}^{w^*}q_{w'_1}P(w'_1,s_1)$ appears because a bubble with at most $w^*$ workers does not choose uniformly among all households, but only among those that also have at most $w^*$ workers.
\end{itemize}

We also compute the fraction of individuals whose bubble size increases after applying this strategy. Since this quantity refers to individuals and not households, each household class must be weighted by its size $s$. Thus,
\begin{eqnarray}
f_{\mathrm{grow}}
&=&
\frac{
\sum_{s=1}^{S_{\max}}\sum_{w=0}^{s}
s\,P(w,s)\,q_w
}{
\sum_{s=1}^{S_{\max}}\sum_{w=0}^{s}
s\,P(w,s)
}
\nonumber\\
&=&
\frac{1}{\langle s\rangle}
\sum_{s=1}^{S_{\max}}\sum_{w=0}^{s}
s\,P(w,s)\,q_w ,
\label{eq.fracPeopleGrowWorkers}
\end{eqnarray}
where $S_{\max}$ is the largest household size in the original household distribution, and $\langle s\rangle=\sum_{s=1}^{S_{\max}}\sum_{w=0}^{s}sP(w,s)=\sum_{s=1}^{S_{\max}}sP(s)$ is the mean household size. It is important to note that Eq.~(\ref{eq.fracPeopleGrowWorkers}) is computed using the original distribution $P(w,s)$, before the merging process is applied. In other words, this calculation identifies which individuals belong to households that are selected to merge by the strategy. The merged distribution $\widetilde{P}(w,s)$ is used later to describe the composition of the bubbles after merging, but it is not used to compute $f_{\mathrm{grow}}$.

The term $sP(w,s)/\langle s\rangle$ in Eq.~(\ref{eq.fracPeopleGrowWorkers}) can be interpreted as the fraction of individuals in the population who live in households of size $s$ with $w$ workers. Multiplying by $q_w$ selects only the households that are allowed to merge. Therefore, Eq.~(\ref{eq.fracPeopleGrowWorkers}) gives the fraction of people whose household belongs to the set of households that  merge and, as a consequence, whose bubble size increases. 

\subsection{Merging strategies based on household size}
This scenario allows households of size $s\le s^*$ to merge with any other household of size $s>s^*$. Let $F=\sum_{s=1}^{s^*}P(s)$ denote the fraction of households of size $s\le s^*$. On the other hand, we define $Q_{s_1,s_2}$ as an indicator function, which is equal to one when simultaneously $s_1 \le s^*$ and $s_2> s^*$, and $Q_{s_1,s_2}=0$ otherwise. The resulting distribution after merging is,
\begin{eqnarray}
\widetilde{P}(w,s)=\frac{1}{c}\left(P(w,s)\frac{1-2F}{1-F}+\sum_{s_1=1}^{s^*}\sum_{w_1=0}^{s_1}P(w_1,s_1)\frac{P(w-w_1,s-s_1)}{1-F}Q_{s_1,s-s_1}\right),
\end{eqnarray}
where,
\begin{itemize}
\item $c$ is a normalization factor to ensure that $\sum_{s=1}^{\infty}\sum_{w=0}^{s}\widetilde{P}(w,s)=1$.
\item the first term on the r.h.s. corresponds to the fraction of households of size $s$ with $w$ workers that did not merge. The factor $(1-2F)/(1-F)$ represents the probability that a household of size $s>s^*$ remains unmerged: of the ($1-F$) fraction of large households, only $(1-2F)$ do not merge with a small household;
\item the second term on the r.h.s. is the fraction of merged bubbles that come from two households: one of size $s_1\le s^*$ with $w_1$ workers, and the other household with size $s-s_1>s^*$ and $w-w_1$ workers. After these two households merge, we obtain a larger bubble with size $s$ and $w$ workers, increasing the value of the probability $\widetilde{P}(w,s)$. The factor $1/(1-F)$ appears because a household of size $s\le s^*$ does not choose uniformly among all households, but only among those with size $s>s^*$, which represent a fraction $1-F$ of all households.
\end{itemize}

We also compute the fraction of individuals whose bubble size increases after applying this household-size-based strategy. As before, this fraction is computed using the original distribution $P(w,s)$, before the merging process is applied. The merged distribution $\widetilde{P}(w,s)$ describes the composition of the bubbles after merging, but it is not used to compute $f_{\mathrm{grow}}$.

In this strategy, there are two contributions to $f_{\mathrm{grow}}$. The first one comes from individuals living in households with size $s\leq s^*$, because these are the households that initiate the merging process. The second one comes from individuals living in larger households, with $s>s^*$, whose household is selected by a smaller household. Thus,
\begin{eqnarray}\label{eq.fracPeopleGrowSize}
f_{\mathrm{grow}}
&=&
\sum_{s=1}^{s^*}\sum_{w=0}^{s}
\frac{sP(w,s)}{\langle s\rangle}
\nonumber\\
&&+
\left[
\sum_{s=1}^{s^*}\sum_{w=0}^{s}
P(w,s)
\right]\cdot
\left[
\sum_{s=s^*+1}^{S_{\max}}\sum_{w=0}^{s}
\frac{sP(w,s)}{\langle s\rangle}\frac{1}{1-F}
\right],
\end{eqnarray}
where $\langle s\rangle=\sum_{s=1}^{S_{\max}}\sum_{w=0}^{s}sP(w,s)$ is the mean household size.

The first term in Eq.~(\ref{eq.fracPeopleGrowSize}) is the fraction of individuals who live, before merging, in households with size $s\leq s^*$. These individuals see their bubble size increase because their household is merged with a larger household. The second term accounts for individuals in households with size $s>s^*$ whose bubble grows because their household is selected by a smaller household. In this term, the first bracket is the fraction of households with size $s\leq s^*$, while the second bracket is the individual-weighted fraction of people living in households with size $s>s^*$. The factor $s/\langle s\rangle$ appears because we are counting individuals, not households. Finally, the factor $1/(1-F)$ appears because households  with size $s\leq s^*$ only select households with size $s> s^*$.

%\appendix
%\counterwithin{figure}{section}
%\counterwithin{table}{section}
%\renewcommand{\thefigure}{\Alph{section}.\arabic{figure}}
%\renewcommand{\thetable}{\Alph{section}.\arabic{table}}
%\pagenumbering{arabic}
%\setcounter{page}{1}
%\setcounter{section}{0}
%\renewcommand{\thesection}{S\arabic{section}}
%\renewcommand{\thesubsection}{\thesection.\arabic{subsection}}
%\setcounter{figure}{0}
%\renewcommand{\thefigure}{S\arabic{figure}}
%\setcounter{equation}{0}
%\renewcommand{\theequation}{S\arabic{equation}}  % Añadido para las ecuaciones
%\setlength{\textwidth}{7cm} 

\bibliography{bib}

\end{document}